\documentclass[pdflatex,sn-mathphys-num]{sn-jnl}

\usepackage{graphicx}%
\usepackage{multirow}%
\usepackage{amsmath,amssymb,amsfonts}%
\usepackage{amsthm}%
\usepackage{mathrsfs}%
\usepackage[title]{appendix}%
\usepackage{xcolor}%
\usepackage{textcomp}%
\usepackage{manyfoot}%
\usepackage{booktabs}%
\usepackage{algorithm}%
\usepackage{algorithmicx}%
\usepackage{algpseudocode}%
\usepackage{listings}%

\theoremstyle{thmstyleone}%
\theoremstyle{thmstyletwo}%

\theoremstyle{thmstylethree}%

\usepackage{booktabs}
\usepackage{amsfonts,amsmath,amssymb}
\usepackage{graphicx,float}
\usepackage{tikz}
\usetikzlibrary{shapes,arrows,positioning}

\usepackage{url,hyperref}
\usepackage{doi}
\usepackage{color,soul}
\usepackage{cleveref}
\crefname{figure}{Fig.\@}{Figs.\@}
\crefname{section}{Sec.\@}{Secs.\@}
\newcommand{\gm}{\mathbf{A}}
\newcommand{\xx}{\mathbf{x}}
\newcommand{\yy}{\mathbf{y}}

\DeclareMathOperator{\pb}{\mathbf{P}\mathopen{}}
\newcommand\Pleft[1]{\pb\mkern-1mu\left[#1\right]}

\begin{document}

\title[Robustness of graph embedding]{Robustness of shallow graph embedding methods for community detection}


\author*[1,2]{\fnm{Zhi-Feng} \sur{Wei}}\email{zfwei@pnnl.gov}

\author[3]{\fnm{Pablo} \sur{Moriano}}\email{moriano@ornl.gov}

\author[3]{\fnm{Ramakrishnan} \sur{Kannan}}\email{kannanr@ornl.gov}

\affil*[1]{\orgdiv{Advanced Computing, Mathematics, and Data Division}, \orgname{Pacific Northwest National Laboratory}, \orgaddress{\city{Richland}, \postcode{99354}, \state{WA}, \country{USA}}}

\affil[2]{\orgdiv{Department of Mathematics}, \orgname{Indiana University}, \orgaddress{\city{Bloomington}, \postcode{47405}, \state{IN}, \country{USA}}}

\affil[3]{\orgdiv{Computer Science and Mathematics Division}, \orgname{Oak Ridge National Laboratory}, \orgaddress{\city{Oak Ridge}, \postcode{37830}, \state{TN}, \country{USA}}}


\abstract{This study investigates the robustness of shallow graph embedding methods for community detection in the face of network perturbations, specifically node deletions. Graph embedding techniques, which represent nodes as low-dimensional vectors, are widely used for various graph machine learning tasks due to their ability to capture structural properties of networks effectively. However, the impact of perturbations on the performance of these methods remains relatively understudied. The research considers state-of-the-art shallow graph embedding methods from two families: matrix factorization (e.g., LE, LLE, HOPE, M-NMF) and random walk-based (e.g., DeepWalk, LINE, node2vec). Through experiments conducted on both synthetic and real-world networks, the study reveals varying degrees of robustness within each family of shallow graph embedding methods. The robustness is found to be influenced by factors such as network size, initial community partition strength, and the type of perturbation. Notably, node2vec and LLE consistently demonstrate higher robustness for community detection across different scenarios, including networks with degree and community size heterogeneity. These findings highlight the importance of selecting an appropriate shallow graph embedding method based on the specific characteristics of the network and the task at hand, particularly in scenarios where robustness to perturbations is crucial.}




\maketitle

\section{Introduction}
The use of low-dimensional vector representations, or graph embeddings, has garnered significant attention across various disciplines~\cite{Goyal-2018, Makarov:2021:Survey:Graph:Embeddings:Applications, Peng:2021:Neural:Embeddings:Periodicals}. These representations play a crucial role in characterizing important functional network properties such as robustness and navigability~\cite{Kleineberg:2017:Geometric:Correlations:Mitigate:Vulnerabilities, Osat:2023:Embedding:Aided:Dismantling,Boguna:2009:Navigability:Networks, Boguna-2010}. Additionally, they are instrumental in performing graph analysis tasks including node classification~\cite{Bhagat:2011:Node:Classification}, link prediction~\cite{Liben:2003:Link:Prediction}, and visualization~\cite{Pereda:2019:Visualization:Graphs}.

Communities, which are sets of nodes more likely to be connected to each other than to the rest of the graph, are fundamental features of many complex systems represented as graphs~\cite{Detection, Fortunato-2016, Moriano:2019:Community:Event:Detection}. Recent research has demonstrated the potential of using graph embeddings for community detection in networks~\cite{Tandon-2021}.  This approach leverages the inherent structural information captured in the vector representations to identify cohesive groups of nodes within the network. Overall, graph embeddings offer a versatile framework for analyzing and understanding complex networks, enabling a wide range of applications across various domains.

Community detection using graph embeddings relies on representing graph nodes as points in a low-dimensional vector space, where communities manifest as clusters of points that are close to each other and sufficiently separated from other clusters~\cite{Tandon-2021, Gu:2021:Selection:Embedding:Dimension, Kojaku:2023:Community:Detection:Neural:Embedding}. These clusters are then recovered using standard data clustering techniques such as $k$-means clustering~\cite{Macqueen:1967:K-Means}. Given the practical importance of consistently detecting communities and the effectiveness of graph embedding approaches in doing so, it is essential to investigate the robustness of graph embeddings for community detection under network perturbations. In this context, \emph{robustness} refers to the ability of a graph embedding method to tolerate perturbations while still accurately recovering the original community structure~\cite{Karrer-2008, Wang-2017, Tian-2023}. This investigation is crucial for understanding how communities identified by graph embedding methods withstand random errors or adversarial attacks, serving as a proxy for evaluating network functionality. Robust graph embeddings ensure the reliability and stability of community detection algorithms in real-world network applications.
Moreover, robustness in graph embeddings has practical implications beyond community detection. 
Robustness to node deletion in community detection reflects a method's ability to recover meaningful structure despite incomplete or perturbed network data. In recommendation systems, such robustness ensures that user or item communities remain stable even when some users or items are absent from the interaction network, thereby supporting consistent and reliable recommendations. In anomaly detection, it enhances the capacity to distinguish between genuine structural anomalies and changes caused by the disappearance or removal of nodes in network communication systems, improving anomaly detection accuracy.

Previous research has investigated the impact of network perturbations on the robustness of community structures using traditional community detection methods~\cite{Wang-2018, Tian-2023}. However, little attention has been given to understanding the robustness of graph embedding methods used for community detection under perturbations in the graph topology. While there are numerous graph embedding techniques available for community detection~\cite{Goyal-2018, Xu-2021}, it remains challenging for practitioners to select a robust method and comprehend its implications across various factors, including the size of the network, strength of the original community structure, type of perturbation, and the specific embedding technique used. Moreover, the majority of research on using graph embeddings for community detection has been conducted on synthetic networks, warranting validation of the effects of perturbations on real-world network data. Addressing these gaps in research, \emph{we aim to evaluate the robustness of graph embedding methods for community detection under various types of graph topology perturbations, and to identify the methods that demonstrate the highest robustness}. By exploring these factors across synthetic and real-world network datasets, we seek to provide insights into the effectiveness and limitations of graph embedding techniques in preserving community structures amidst network perturbations.

In this paper, we address this gap of knowledge by devising a systematic framework to investigate the robustness of graph embedding methods used to extract community structures under network perturbations. 
In particular, we focus on perturbations induced by node deletion, which models the disruption or loss of entities together with their connections in a network. Other perturbation mechanisms, such as independent edge removal or the introduction of spurious edges due to modeling noise, represent alternative scenarios that may affect embedding behavior differently.
We consider both synthetic Lancichinetti-Fortunato-Radicchi (LFR) benchmark graphs~\cite{LFR1} and empirical networks with degree and community size heterogeneity. 
The LFR benchmark is particularly suitable for robustness experiments because it allows explicit control over the mixing parameter that governs the clarity of the planted community structure while preserving heterogeneous degree and community size distributions.
The LFR benchmark graphs provide a controlled environment in which network properties such as community strength, size, and degree heterogeneity can be systematically varied, while the empirical networks serve as representative case studies that allow us to examine whether the observed robustness trends also appear in real-world systems.

Our perturbation approach involves the removal of nodes, where nodes are selected either randomly ~\cite{Wang-2018} or through targeted selection using betweenness centrality~\cite{Albert-2000, Cohen:2000:Resilience:Internet:Random:Breakdowns, Cohen:2001:Breakdown:Internet:Intentional:Attack}. When a node is removed, all of its incident edges are removed as well, resulting in a structural perturbation of the network.
We use seven state-of-the-art shallow graph embedding methods, namely Laplacian eigenmap (LE)~\cite{Belkin-2003}, locally linear embedding (LLE)~\cite{Sam-2000}, higher-order preserving embedding (HOPE)~\cite{Ou-2016}, modularized non-negative matrix factorization (M-NMF)~\cite{WangX-2017} in the matrix factorization family; and DeepWalk~\cite{Perozzi-2014}, large-scale information network (LINE)~\cite{Tang-2015}, and node2vec~\cite{Grover-2016} in the random walk family. We use $k$-means based on spherical distance~\cite{Schubert-2021} to achieve more effective clustering results in high-dimensional spaces. 
These clusters are then recovered using standard data clustering techniques such as $k$-means clustering~\cite{Macqueen:1967:K-Means}. In this work, the embedding–clustering pipeline is used as an analytical framework to evaluate how structural perturbations affect the community structure encoded by shallow graph embedding representations, rather than to benchmark embedding methods against traditional topology-based community detection algorithms.
We use element-centric similarity (ECS)~\cite{ECS}, a widely adopted community similarity metric, to quantify changes in the community structure after 
perturbations as a proxy of characterizing graph embedding robustness. This comprehensive methodology allows us to systematically assess and compare the impact of various perturbation strategies on the robustness of graph embedding methods for community detection in both synthetic and real-world networks. 

Our findings indicate that incremental removal of nodes reduces ECS. 
However, this impact tends to be more noticeable across shallow graph embedding methods, but node2vec and LLE, and varies across network sizes, the initial community structure strength, and the perturbation type.
This aligns with the expectation that removing a greater number of nodes results in a more substantial disruption of the community structure. Notably, targeted node removal leads to a more pronounced and accelerated decline in ECS compared to random node removal. This is rationalized by the fact that adversarial attacks tend to swiftly dismantle the network's community structure. Among the seven shallow network embedding methods considered, node2vec and LLE consistently prove to be more robust for community detection within their respective graph embedding families; these two methods provide the highest ECS similarity scores. Overall, node2vec is the most robust shallow graph embedding method for community detection. Our results generally concur with those of Kojaku et al.~\cite{Kojaku:2023:Community:Detection:Neural:Embedding}, which found that node2vec learns more consistent community structures with networks of heterogeneous degree distributions and size. 

We notice that community detection methods based on graph neural networks (GNNs) currently constitute a popular approach \cite{su-2024}. The focus of our study, however, is on shallow graph embedding methods rather than GNNs.
Recent research in graph representation learning has increasingly explored attribute-aware embeddings, particularly through GNNs, which integrate node features with network topology. In this work, however, we focus on topology-only networks without node attributes and investigate the robustness of shallow structural embedding methods under structural perturbations. Such topology-based embeddings remain widely used due to their conceptual simplicity, scalability, and applicability to networks where attribute information is unavailable. Accordingly, our analysis concentrates on widely used topology-based embeddings that rely solely on network structure. Robustness of GNN-based community detection under both structural and attribute perturbations has been studied separately in a companion work by the authors~\cite{Goel:2025:Community:Detection:Robustness:GNN}, which evaluates several GNN architectures under feature noise, topology distortions, and adversarial attacks.

We have made available the code and data to reproduce all the results at 
\footnote{\url{https://t.ly/uIclh} (accessed on August 1, 2025)},
\footnote{\url{https://t.ly/OzFzV} (accessed on August 1, 2025)}. 

\section{Related Work}
\subsection{Shallow graph embedding for community detection}
There has been substantial research in network community detection leveraging shallow graph embedding techniques.

Tandon et al.\ \cite{Tandon-2021} tested the ability of the aforementioned seven state-of-the-art graph embedding techniques to detect communities on benchmark graphs. It was reported that if the parameters of the embedding techniques are suitably chosen, the performance of community detection methods based on graph embedding techniques is comparable to the traditional community detection algorithms, including Infomap and Louvain.

Zhang et al.\ \cite{LS1-4} proposed a method that simultaneously performs network embedding and community detection in directed networks. Their model introduced separate vector representations for out-nodes and in-nodes, allowing nodes to belong to different out- and in-communities. Their method has improved performance in community detection while jointly estimating node embeddings and their community assignments. An efficient alternating optimization scheme was developed, and theoretical guarantees for both embedding and community detection are provided, supported by experiments on simulated and real-world networks.

Ghanbari et al.\ \cite{LS1-6} investigated the efficiency of different node embedding algorithms combined with clustering methods for community detection. They evaluated four embedding techniques, including leveraging convolution, attention, inductivity, and shallowness, paired with three clustering algorithms, including KMeans++, DBSCAN, and hierarchical clustering. Experimental results showed that GraphSAGE, an inductive embedding method, combined with KMeans++ produced the best performance. 

Zhu et al.\ \cite{zhu-1} proposed a structural equivalence embedding method based on non-negative matrix factorization (SENMF) for community detection, which embeds two similarity measures one capturing node closeness and the other reflecting similarity between distant nodes into a low-dimensional space. Experimental results show that SENMF outperforms several well-known network embedding and traditional community detection methods. In fact, as mentioned in \cite{zhu-1}, when taking parameters $\theta=0$ and $\varphi=0$, SENMF will equal M-NMF.

Kovács et al.\ \cite{LS1-1} proposed an iterative embedding and reweighting procedure that strengthens intra-community links while weakening inter-community links, enhancing cluster visibility and detectability. This process can lead to such strong geometric separation that even simple link parsing recovers communities with high precision. Additionally, it can serve as a pre-processing step to improve traditional community detection algorithms, including Louvain, Infomap, and HDBSCAN.

Pankratz et al.\ \cite{LS1-5} proposed a method to enhance community detection by initializing heuristic algorithms with partitions obtained from network embeddings. Their approach embeds nodes into a latent space to filter noise while preserving structural relationships, followed by clustering to generate a stable partition. This initialization improves the performance and reduces the volatility of traditional methods such as Louvain and Leiden, particularly in networks with weak community structures. Experimental results demonstrated that the effectiveness of this procedure depends on network topology rather than size.

Zhu et al.\ \cite{zhu-2} proposed a novel Graph Convolutional Network (GCN) method for unsupervised community detection, termed UCDMI. This method leverages mutual information maximization to capture fine-grained information about the global network structure without supervision. A new aggregation function for GCN is introduced to differentiate the importance of neighboring nodes, enhancing the quality of node representations. The proposed approach demonstrates significant improvements in community detection performance through extensive experiments, outperforming several state-of-the-art methods. This method offers a promising direction for unsupervised community detection by integrating mutual information maximization with GCNs.

In this work, we focus on community detection in undirected networks using the seven graph embedding techniques from \cite{Tandon-2021}. Rather than introducing new community detection methods, we examine the robustness of these embedding techniques under network perturbations.

\subsection{Robustness of community detection methods under network perturbations}
The following articles examines the robustness of community structures using traditional community detection methods under different types of network perturbations.

Karrer et al.\ \cite{Karrer-2008} proposed a framework for evaluating the significance of community structures by perturbing networks with configuration model random graphs and measuring the resulting changes using variation of information. They employed a spectral optimization method for modularity to perform community detection. This approach was applied to a wide range of networks, including both real-world and synthetic examples.

Ramirez-Marquez et al.\ \cite{LS2-1} analyzed the effects of weight uncertainties on community structures in networks with weighted edges. Using Monte Carlo-based uncertainty propagation, they analyzed how variations in link weights influence community structures. The approach was demonstrated through the analysis of an electric power system network, highlighting the significance of weight uncertainties in determining reliable community partitions.

Peche et al.\ \cite{LS2-2} investigated perturbations where the connections between network nodes are influenced by a random geometric graph. They demonstrated that community detection using spectral methods remains robust to such noise, based on a thorough and detailed analysis of the spectrum of the random graph's adjacency matrix.

Jia et al.\ \cite{LS2-3} proposed a robustness guarantee for community detection against adversarial structural perturbations. For any given community detection method, they introduced a smoothed community detection approach by randomly perturbing the graph structure. They theoretically demonstrated that 
the smoothed community detection method provably groups a given arbitrary set of nodes into the same community (or different communities) provided that the number of edges added or removed by an attacker is bounded. The method was also empirically evaluated on multiple real-world graphs with ground truth communities.

Tian et al.\ \cite{Tian-2023} studied the robustness of community detection algorithms under edge addition perturbations in synthetic and empirical temporal networks. They considered two edge addition scenarios---random and targeted---and evaluated the robustness of communities detected by Infomap, Label Propagation, Leiden, and Louvain algorithms. It was found that robustness performance varies across different community similarity metrics and detection methods, highlighting the dependence of community stability on the choice of algorithm.

\subsection{GNNs for community detection}
Furthermore, given that graph neural network-based community detection methods are numerous and currently constitute a popular approach, to reflect the current research trends, we include the following works on graph community detection using graph neural networks.

Zhang et al.\ \cite{zhu-2021} proposed a deep learning-based approach for dynamic community detection that incorporates node relevancy to capture structural information. Additionally, it employs cross-entropy to smooth learned features between adjacent time steps, addressing potential large discrepancies that don't align with real-world dynamics.

Sun et al.\ \cite{sun-2022} introduced a novel graph neural network encoding method for multiobjective evolutionary algorithms to address community detection in complex attribute networks. The approach transforms the problem into one with continuous decision variables, facilitating smoother fitness landscapes, and demonstrates superior performance on both single- and multi-attribute networks.

Tsitsulin et al.\ \cite{JMLR:v24:20-998} developed Deep Modularity Networks (DMoN), an unsupervised pooling method inspired by the modularity measure of clustering quality. Their work addresses the critical observation that existing GNN pooling methods are surprisingly ineffective for graph clustering. On real-world data, they demonstrated that DMoN produces high-quality clusters strongly correlated with ground truth labels, achieving state-of-the-art results with over $40\%$ improvement over other pooling methods.

Su et al.\ \cite{su-2024} provided a comprehensive survey of recent advancements in deep learning techniques for community detection, particularly highlighting their utility in handling high-dimensional network data. It proposes a new taxonomy that categorizes state-of-the-art methods, including various deep neural network models, and summarizes benchmark datasets, evaluation metrics, and open-source implementations.

To address the limitations of previous research, such as the lack of a systematic evaluation of perturbation impact on community structure robustness, our study examines node deletion based on both random and targeted node removal. We apply this approach to both synthetic and real-world networks, accounting for degree and community size heterogeneity. This perturbation approach presents a novel method of network perturbation with real-world relevance. To compare the community structures before and after the network perturbation, we utilized ECS, a robust metric for evaluating network clusters.

\section{Methods}

\subsection{Shallow Graph Embedding Methods}

In this part, we introduce the shallow graph embedding methods we used in our study.

We use $\mathbf{A}$ to denote the adjacency matrix of a graph.
The graph embedding methods are classified into two families (i.e., matrix-based family in \Cref{sec:mf} and random walk family in \Cref{sec:rw}).

\subsubsection{Matrix Factorization Methods}\label{sec:mf}
Matrix-based methods project network nodes into a Euclidean space via eigenvectors following a similar idea as in spectral clustering \cite{LUX}.
In this family of methods, we include Laplacian eigenmap (LE) \cite{Goyal-2018,Belkin-2003}, locally linear embedding (LLE) \cite{Tandon-2021, Sam-2000}, higher-order preserving embedding (HOPE) \cite{Goyal-2018,Ou-2016}, and modularized non-negative matrix factorization (M-NMF) \cite{Tandon-2021,Wang-2017}.

\paragraph{Laplacian Eigenmap (LE)}
Laplacian eigenmap (LE) \cite{Goyal-2018,Belkin-2003} aims to minimize the objective function
\begin{displaymath} 
\sum_{i,j}
\left| \xx_i - \xx_j \right|^2\gm_{ij},
\end{displaymath}
subject to the constraint $\xx^T\mathbf{D}\xx=\mathbf{I}$, where $\mathbf{D}$ represents the diagonal matrix of graph node degrees, and
$\xx_i$ denotes the vector indicating the position of the point representing node $i$ in the embedding.
For $d$-dimensional embedding with LE, the solution can be obtained by extracting the eigenvectors corresponding to the $d$ smallest eigenvalues (except the zero eigenvalue) of the normalized Laplacian matrix 
\begin{displaymath} 
\mathbf{D}^{-1}(\mathbf{D}-\gm).
\end{displaymath}

\paragraph{Locally Linear Embedding (LLE)}
Locally linear embedding (LLE) \cite{Tandon-2021, Sam-2000} minimizes the objective function
\begin{displaymath} 
\sum_i
\Bigl|\xx_i-\sum_j\gm_{ij}\xx_j\Bigr|^2.
\end{displaymath}
Given the form of the objective function, each point in the embedded space $\xx_i$ is approximated as a linear combination of its neighbors in the original graph.
To ensure a well-posed problem, solutions are required to be centered at the origin, i.e., $\sum_i \xx_i= 0$, and to have unit variance, i.e., $\frac{1}{N}\xx^T\xx = \mathbf{I}$.
Subject to these constraints, for a $d$-dimensional embedding with LLE, the solution is approximated by the eigenvectors corresponding to the lowest $d$ eigenvalues (excluding the zero eigenvalue) of the matrix $(\mathbf{I}-\gm)^T(\mathbf{I}-\gm)$.

\paragraph{Higher-Order Preserving Embedding (HOPE)}
Higher-order preserving embedding (HOPE) is designed to maintain the similarity between nodes \cite{Goyal-2018,Ou-2016}. The objective function being minimized is $\bigl|\mathbf{S} - \xx\xx^T\bigr|$, where $\mathbf{S}$ is the Katz similarity matrix defined as 
\begin{displaymath}
\mathbf{S} = \beta\sum_{\ell=1}^{\infty}\gm^\ell,
\end{displaymath} 
with $\beta$ denoting the decay parameter. In our experiments, we set $\beta=0.01$.

\paragraph{Modularized Non-Negative Matrix Factorization (M-NMF)}
Modularized Non-Negative Matrix Factorization (M-NMF) \cite{Wang-2017} preserves both the first-order proximity (pairwise node similarity) and the community structure for network embedding. Modularity, which evaluates the quality of community partitions within a network by comparing the observed network with randomized versions that lack inherent community structure, 
is integrated into the optimization function of M-NMF.
While minimizing the optimization function of M-NMF (see \cite{Wang-2017}), nodes are brought into proximity when they exhibit similarity and simultaneously when they are part of clusters derived from high-modularity partitions \cite{Tandon-2021}. 

The following parameter values are used in our experiments: $\lambda=0.2$, $\alpha=0.05$, $\beta=0.05$, $\eta=5.0$, and number of iterations $N=200$.

\subsubsection{Random Walk Methods}\label{sec:rw}
Random walk methods learn network embeddings of graph nodes by modeling a stream of short random walks.
We chose DeepWalk \cite{Perozzi-2014}, LINE \cite{Tang-2015}, and node2vec \cite{Grover-2016} as widely used methods of this family.

\paragraph{DeepWalk}
DeepWalk \cite{Perozzi-2014} extends language modeling techniques to graphs, departing from words and sentences. This algorithm leverages local information acquired through random walks, treating these walks as analogous to sentences in the word2vec \cite{Mikolov-2013} language modeling approach.
To generate a random walk originating from a specified starting node, neighbors of the current node in the walk are randomly selected and added to the walk iteratively until the intended walk length is achieved.

In our experiments, we use the following parameters: random walk length $t=40$, window size $w=10$, number of walks per node $n=80$.

\paragraph{Large-Scale Information Network Embedding (LINE)}
The large-scale information network embedding (LINE) method aims to position nodes in close proximity to each other as their similarity increases. LINE can be implemented based on first-order similarity, second-order similarity, or both. According to \cite{Tandon-2021}, it works the best when implemented based on first-order similarity, i.e., adjacency of nodes. In fact, for unweighted networks, the LINE method based on first-order similarity minimizes this objective function:
$$
-\sum_{(i,j)\in E}\log\Bigl(\frac{1}{1+\exp{(-\xx_i\cdot\xx_j)}}\Bigr),
$$
where $E$ is the edge set and $\xx_i$ denotes the embedding point of node $i$.

The optimization process of the LINE method utilizes stochastic gradient descent, which is enhanced by an edge-sampling treatment, as detailed in \cite{Tang-2015}. Specifically, the utilization of the LINE embedding method necessitates the presence of GPU for computation. In our study, the machine learning model was trained using a batch size of $8192$ over $100$ epochs, facilitating iterative updates to the embedding vector across the entire training dataset.

\paragraph{node2vec}
node2vec employs a similar optimization procedure to DeepWalk, but the process of generating ``sentences'' differs \cite{Grover-2016}. Specifically, simple random walk is used for DeepWalk, while biased random walk is utilized for node2vec. These biased random walks are composed of a blend of steps following both breadth-first and depth-first search strategies, with parameters $p$ and $q$ controlling the respective influences of these two strategies (see \cite{Grover-2016} for details about how random walk are generated in node2vec).
Therefore, node2vec produces higher-quality and more informative embeddings than DeepWalk. 

We use the following parameters in our experiments: random walk length $t=10$, window size $w=10$, number of walks per node $n=80$, biased walk weights $p=1$ and $q=1$.

\Cref{tab:table2} summarizes the experimental parameters used by the embedding methods, excluding the embedding dimension $d$. The LE and LLE methods require only the embedding dimension $d$, while the remaining methods include additional parameters. For these parameters, we adopt the default settings reported in Tandon et al. (2021)~\cite{Tandon-2021}, who show that the default configurations yield performance comparable to traditional community detection methods and that parameter optimization provides only limited additional improvement.

\begin{table}[htbp]
\caption{\label{tab:table2} Parameter values for embedding methods.}
\centering
\begin{tabular}{@{}p{0.23\linewidth} p{0.17\linewidth} p{0.50\linewidth}@{}}
\toprule
\textbf{Family} & \textbf{Method} & \textbf{Parameters} \\
\midrule
Matrix         & HOPE            & $\beta=0.01$ \\
Matrix         & M-NMF           & $\lambda=0.2$, $\alpha=\beta=0.05$,\\
               &                  &$\eta=5$, $N=200$ \\
Random Walk    & DeepWalk        & $t=40$, $w=10$, $n=80$ \\
Random Walk    & LINE            & batch size: $8192$, epochs: $100$ \\
Random Walk    & node2vec        & $t=10$, $w=10$, $n=80$,\\
                &                   &$p=q=1$ \\
\bottomrule
\end{tabular}
\end{table}
\subsection{LFR Benchmark}\label{sec:LFR}
Lancichinetti--Fortunato--Radicchi (LFR) benchmark graphs \cite{LFR1,LFR2} have become a widely adopted tool for generating graphs with ground-truth communities. One of the primary advantages of utilizing the LFR benchmark is its ability to produce graphs where both degree and community size distributions follow power-law distributions. These distributions closely resemble the properties observed in many real-world networks \cite{LFR1,Clauset-2004,Guimera-2003,Palla-2005}. The exponents governing the degree and community size distributions are controlled by the parameters $\alpha$ and $\beta$, respectively.
To generate LFR benchmark graphs, one must specify various other parameters, including the average degree $\langle k \rangle$, maximum degree $maxk$, 
maximum community size $maxc$, and the mixing parameter $\mu$. The mixing parameter $\mu$ represents the fraction of nodes that share edges across different communities, with lower values of $\mu$ leading to a higher ratio between internal and external edges. This results in increased modularity, a commonly used quality metric for assessing community strength in community research \cite{Detection}. Thus, smaller values of $\mu$ correspond to stronger partitions. 
The range $\mu \in \{0.01, 0.1, 0.2, 0.3, 0.4, 0.5\}$ spans community structures from well separated to highly diffuse. At low values of $\mu$, most edges connect nodes within the same community, resulting in clearly defined communities. As $\mu$ increases, an increasing fraction of edges connects different communities, making community boundaries progressively weaker and community detection more challenging. Although the mixing parameter does not alter the degree or community size distributions of the generated networks, it enables a systematic evaluation of embedding robustness under increasingly weak community organization.
Note that LFR benchmark graphs are undirected and unweighted.

We generate LFR benchmark graphs using the \texttt{LFR\_benchmark\_graph} method in the publicly available Python package \texttt{NetworkX} \cite{nx}. Our experiments involve two different network sizes: $1,000$ nodes and $10,000$ nodes, allowing us to investigate the impact of perturbations at varying scales. Furthermore, we explore the influence of the initial community strength by adjusting the mixing parameter, $\mu$. The specific parameter values utilized for the generation of LFR benchmark graphs are detailed in Table~\ref{tab:table1}. Note that the proposed experimental design can be applied to any other synthetic graph generators such as the Girvan--Newman benchmark \cite{GN} and the stochastic block model \cite{SB}.

\begin{table*}[htbp]
\caption{\label{tab:table1}%
Synthetic experiment parameters.}
\centering
\begin{tabular}{@{}clc@{}} 
\toprule
\textbf{Parameter} & \textbf{Description} & \textbf{Value} \\ 
\midrule
$N$ & number of nodes & $1,000$ and $10,000$ \\ 
$\texttt{maxk}$ & max degree for LFR & $0.1 N$ \\ 
$\langle k \rangle$ & average degree for LFR & $25$ \\ 
$\texttt{maxc}$ & max community size for LFR & $0.1 N$ \\ 
$\texttt{minc}$ & min community size for LFR & $50$ \\ 
$\alpha$ & degree distribution exponent for LFR & $-2$ \\ 
$\beta$ & community size distribution exponent for LFR & $-1.1$ \\ 
$\mu$ & mixing parameter for LFR & $0.01, 0.1, 0.2, 0.3, 0.4, 0.5$ \\ 
$r$ & number of realizations & $50$ \\ 
\bottomrule
\end{tabular}
\end{table*}

\subsection{Node Removal Procedure}\label{sec:pert}
In this study, the perturbation process corresponds to node deletion, where a selected node is removed together with all of its incident edges. This procedure models the removal of entities in a network and the associated loss of their connections.
Our node removal procedure investigates the impact of eliminating nodes in the network structure. We employ two distinct methods: node removal based on random node selection and node removal based on targeted node selection. 

Let's first discuss node removal based on random node selection. This simulates random network errors in real-world scenarios~\cite{Rosenkrantz:2009:Resilience:Metrics:Service:Networks, Cheng:2010:Bridgeness:Edge:Significance}. We do so by selecting $p\%$ of nodes randomly and deleting them.
To make our experiment unbiased, we independently generate $50$ distinct sets of nodes. In other words, the number of realizations is $r=50$ (see Table~\ref{tab:table1}).
We ensure that the removal of each set of nodes does not disconnect the remaining part of the network, which is required by the graph embedding methods.
In the event that the remaining part of the network becomes disconnected after the removal of a set of nodes, we omit this set of nodes and proceed with re-sampling.
After each removal, we measure the similarity between the community structure of the remaining network calculated using data clustering in the embedding space (i.e., $k$-means; see more details in \Cref{sec:commdetec}) and the community structure of the original LFR network using community similarity metrics (i.e., ECS \cite{ECS}) as detailed in \Cref{sec:simi}. The overall similarity score for a given $p\%$ value is calculated by averaging the similarity scores across all $50$ sets. 

Second, we discuss targeted node removal. Targeted node removal is based on the betweenness centrality of nodes. Betweenness centrality measures the importance of a node in a network based on how often it lies on the shortest paths between other nodes \cite{BTWN}.
Nodes with high betweenness centrality usually indicate nodes of vital importance to the network, so removing these nodes might cripple the functionality of the network \cite{Bellingeri-2020}. This approach models an adversarial attack in real-world situations~\cite{Zeng:2012:Enhancing:Robustness:Malicious:Attacks, Kocc:2014:Impact:Cascading:Failures:Power:Grid}. To implement node removal based on targeted node selection, we first rank the nodes in the network by their betweenness centrality. The node with the lowest betweenness centrality has rank $1$, and the node with the highest betweenness centrality has rank $N$, where $N$ is the number of nodes in the network. 
We then independently generate $50$ sets of nodes, each containing $p\%$ of nodes in the network. The nodes in each set are selected by randomly sampling nodes with probability proportional to their betweenness centrality rank. We ensure that the removal of each set of nodes does not disconnect the remaining part of the network, which is required by graph embedding methods. 
In the event that the remaining part of the network becomes disconnected after the removal of a set of nodes, we omit the set of nodes and proceed with re-sampling.
After each removal, we measure the similarity between the community structure of remaining network computed through data clustering in the embedding space (i.e., $k$-means; see more details in \Cref{sec:commdetec})
and the ground-truth community structure of the original LFR network using ECS. The overall similarity score for a given $p\%$ is calculated by averaging community similarity scores across all $50$ sets.

In our experiments, the proportion of removed nodes starts from $5\%$ with an increment of $5\%$. The upper bound of proportion 
depends on the network structure and the node removal method. For example, when we have a network with $N=1,000$, $\mu=0.01$, and use node removal based on targeted node removal, it is difficult to keep the network connected when deleting $35\%$ of nodes. In these cases, we use a proportion of removed nodes that is below the threshold, e.g., $5\%$, $10\%$, $15\%$, $20\%$, $25\%$, and $30\%$.
In addition, we also perform data clustering on the embeddings of the original LFR networks with no node removed and compare the results with the ground-truth community structure. This result is recorded as ``0\% of nodes removed.'' We also captured the variation of the measurements by computing standard deviations. 
We compute both the mean ECS values and their standard deviations across multiple realizations of the perturbation process for each experimental configuration. In all cases, the observed standard deviations are very small relative to the corresponding mean values (typically in the range 0–0.05, compared to mean ECS values between 0.5 and 1.0). These small variations do not affect the relative trends among embedding methods, as verified in preliminary plots with error bars. For clarity of visualization, we therefore only report the mean ECS curves in the figures.

\subsection{Data Clustering}\label{sec:commdetec}
Community detection (or clustering) is the task of grouping nodes in a network based on their connectivity. A general consensus is that nodes within the same community are more densely connected than nodes across different communities. There are many different community detection algorithms, each with its own strengths and weaknesses \cite{Tandon-2021}. The best algorithm for a particular network depends on the properties of the network and performance constraints \cite{HHH}.

In this article, we perform community detection based on a two-step process consisting of shallow graph embedding \cite{Xu-2021} and the $k$-means algorithm \cite{Lloyd-1982} for data clustering. That is, we control the data clustering step by using the $k$-means method to optimally find the clusters in the low-dimensional representation. Specifically, we perform shallow graph embedding to assign a point in the Euclidean space to each node in the network. Here, we focus on embedding each node of the graph rather than the whole graph (more details on node-embedding methods can be found in \cite{Yan-2007}). We use seven state-of-the-art shallow graph embedding methods studied in previous research \cite{Tandon-2021} across two different families of graph embedding methods, including: (1) matrix-based methods and (2) random walk methods (according to the taxonomy by M.\ Xu in \cite{Xu-2021}). In the matrix-based family, we include Laplacian eigenmap (LE) \cite{Belkin-2003}, locally linear embedding (LLE) \cite{Sam-2000}, higher-order preserving embedding (HOPE) \cite{Ou-2016}, and modularized non-negative matrix factorization (M-NMF) \cite{Wang-2017}. In the random walk family, we include DeepWalk \cite{Perozzi-2014}, large-scale information network (LINE) \cite{Tang-2015}, and node2vec \cite{Grover-2016}. 
Note that we make the distinction between graph embedding families to stress the foundational mechanism under which graph embeddings are constructed. We, however, note that under specific conditions regarding the window size of the skip-gram model, random walk methods are equivalent to matrix factorization methods \cite{Qiu-2018}. 
Tandon et.\ al provide a thorough test on shallow graph embedding methods for data clustering, which have been shown to perform well for community detection in networks \cite{Tandon-2021}. 

We use the publicly available Python package \texttt{nxt\_gem} for LE and HOPE \cite{nxt-gem}. As there is no correct implementation of LLE for graphs available online, we developed our own implementation available on GitHub 
in \footnote{{h}ttps://github.com/zf-wei/LLECupy (accessed on April 29, 2024)}.
The package \texttt{karateclub} includes implementations of DeepWalk and M-NMF \cite{karate}. The package \texttt{GraphEmbedding} for LINE is publicly available on GitHub in \footnote{{h}ttps://github.com/shenweichen/GraphEmbedding (accessed on April 29, 2024)}. We utilize \texttt{node2vec} package for node2vec embedding available online \footnote{{h}ttps://pypi.org/project/node2vec (accessed on April 29, 2024)}.

After computing graph embeddings, we use the $k$-means algorithm to group the points in the Euclidean space into clusters. 
Note that other data clustering algorithms used before to cluster embedding results, such as Gaussian mixture models \cite{Reynolds-2009}, lead to similar results as $k$-means \cite{Tandon-2021}.
The $k$-means algorithm, a popular method in data clustering, minimizes the distance between each data point and its centroid. It works by iteratively assigning data points to the nearest cluster centroid and then updating the centroids based on the average of the assigned points until convergence, aiming to minimize the within-cluster variance. In our case, the data points are the embedding points in Euclidean space for each network node and centroids are virtual points representing the clusters.
As in high-dimensional spaces, the concept of proximity may not be meaningful as different distances may select different neighbors for the same point \cite{Bayer-1999}; we use $k$-means based on spherical distance to get better clustering outcome. According to \cite{Schubert-2021}, $k$-means based on spherical distance is a widely used clustering algorithm for sparse and high-dimensional data. Specifically, for two nodes embedded as points $\xx=(x_1,x_2,\ldots, x_d)$ and $\yy=(y_1,y_2,\ldots, y_d)$ in the $d$-dimensional Euclidean space, 
the spherical distance between the two nodes is defined as
\begin{displaymath}
\mathsf{d}_S(\xx,\yy)=
1-\frac{\sum_{i=1}^{d}x_iy_i}{\sqrt{\sum_{i=1}^{d}x_i^2\sum_{i=1}^{d}y_i^2}}.
\end{displaymath}

As data clustering methods often necessitate prior knowledge of the number of clusters, we provide the data clustering procedure with the correct number of clusters derived from the LFR benchmark graphs. By doing so, we ensure a fair evaluation of the embedding methods. Note that such a luxury may be absent when analyzing real-world networks, as the true number of clusters is typically unknown in those cases.

\subsection{Community Similarity}
\label{sec:simi}
To measure the similarity between community partitions, we employ ECS \cite{ECS},
computed with the \texttt{CluSim} package \cite{clusim}.  
Compared to other clustering similarity metrics such as the normalized mutual information (NMI) \cite{NMI}, ECS is a more robust metric for comparing clusters and better addresses challenges such as biases in randomized membership, skewed cluster sizes, and the problem of matching \cite{ECS}. We use default parameter values in ECS computation.
ECS yields values in the range from $0$ to $1$, where a higher value indicates a greater similarity between partitions. 

\Cref{scheme} is a schematic diagram showing the process we follow in our experimental design.

\begin{figure}[H]
\centering
\begin{tikzpicture}[node distance=0.6cm, auto]

\tikzset{
    rect/.style={rectangle, draw, rounded corners, text centered, minimum height=1cm, minimum width=2cm, thick},
    arrow/.style={->, >=stealth, thick},
}

\node (lfr) [rect] {Network Generation, \Cref{sec:LFR}};
\node (perturbation) [rect, below=of lfr] {Perturbation, \Cref{sec:pert}};
\node (embedding) [rect, right=of perturbation] {Data Clustering, \Cref{sec:commdetec}};
\node (similarity) [rect, above=of embedding] {Similarity, \Cref{sec:simi}};

\draw [arrow] (lfr) -- (perturbation);
\draw [arrow] (perturbation) -- (embedding);
\draw [arrow] (embedding) -- (similarity);

\end{tikzpicture}
\caption{Schematic diagram of experimental design.}
\label{scheme}
\end{figure}

\section{Results and Discussion}\label{sec:result}
In this section, we present the results of our experiments. 

In the subsequent line figures, unless otherwise stated (in \Cref{986s,23748s,line_high}), subfigures in the top row, i.e., (a) and (b), correspond to random node removal. More specifically, subfigure (a) corresponds to experimental results on small networks with $1,000$ nodes and subfigure (b) depicts experimental results on large networks with $10,000$ nodes. 
Similarly, subfigures in the bottom row, i.e., (c) and (d), correspond to targeted node removal, for small and large networks respectively.

We present line figures using $16$-dimensional and $32$-dimensional embedding. We focus only in these two dimensions as smaller embedding dimensions seem to be enough to provide an accurate low-dimensional vector representation of networks~\cite{Gu:2021:Selection:Embedding:Dimension, Kojaku:2023:Community:Detection:Neural:Embedding}.
Recall that a higher ECS value indicates a greater similarity between network community partitions obtained from the LFR benchmark graphs and the embeddings. Thus, when analyzing line figures, higher ECS values and therefore higher positions of line curves indicate a more robust graph embedding method for community detection.
The following analyses examine how the robustness behavior of shallow graph embedding methods evolves under different experimental factors. These figures are intended to illustrate robustness trends and comparative patterns across the considered perturbation settings.

Finally, \Cref{sec:real} presents experimental results to evaluate the robustness of our community detection methods on two real-world networks with labeled communities: the email-EU-core network \cite{Jure-2007} and the AS network \cite{Boguna-2010}. 
These networks exhibit similar characteristics to synthetic counterparts, specifically power-law degree and community size distributions as observed in LFR benchmark graphs. Our experimental methodology and insights for real-world networks mirror findings for synthetic counterparts.

\subsection{Matrix Factorization Methods}
\subsubsection{Laplacian Eigenmap (LE)}
\begin{figure}[htbp] \centering
\includegraphics[width=8.3cm]{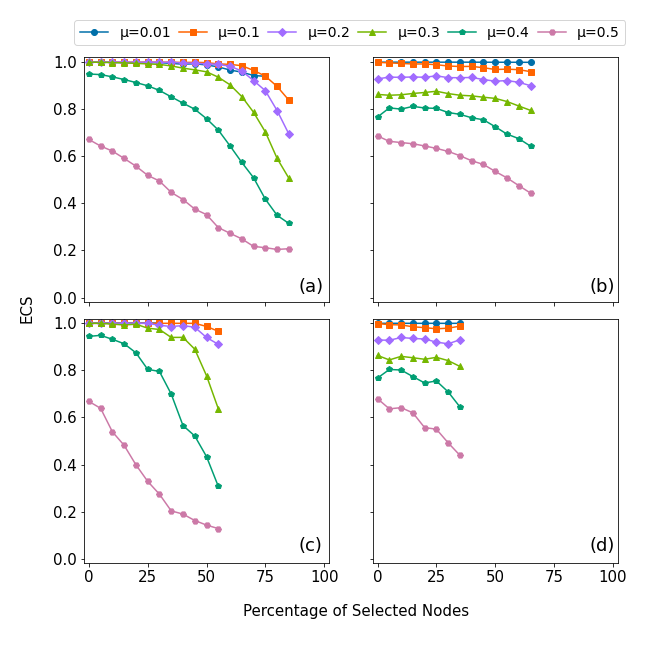}
\caption{Mean ECS using LE method over the percentage of removed nodes on LFR benchmark graphs with $16$-dimensional embeddings. Random node removal is used in (a) and (b), while targeted removal is used in (c) and (d). (a) and (c) are for networks with $1,000$ nodes, and (b) and (d) are for networks with $10,000$ nodes.}
\label{2sk}
\end{figure}

\begin{figure}[htbp] \centering
\includegraphics[width=8.3cm]{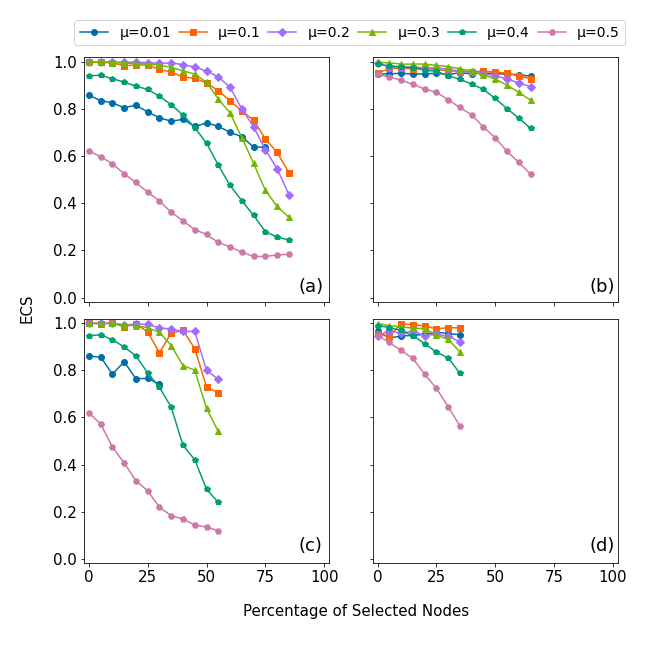}
\caption{Mean ECS using LE method over the percentage of removed nodes on LFR benchmark graphs with $32$-dimensional embeddings. Random node removal is used in (a) and (b), while targeted removal is used in (c) and (d). (a) and (c) are for networks with $1,000$ nodes, and (b) and (d) are for networks with $10,000$ nodes.}
\label{2sw}
\end{figure}

Our experimental results on the impact of node removal, generated through the LE embedding method with dimension $16$, are reported in \Cref{2sk}. 
Our experimental results using LE embedding method with dimension $32$ are collected in \Cref{2sw}.

In \Cref{2sk,2sw}, the experimental results on the impact of node removal generated by LE are demonstrated. The effects of the initial community partition's strength (with stronger partitions experiencing less perturbation impact, i.e., lower $\mu$ values) and the type of perturbation (targeted node removal having a significant influence on clusters) are observed.
Specifically, for networks with strong community partition (i.e., $\mu=0.01$ or $\mu=0.1$), we notice that embedding dimension tends to have a stronger effect on the decay rate of similarity, despite node removal strategy. 
For instance, in \Cref{2sk}(a), with an embedding dimension of $16$ under random node removal, the curve corresponding to $\mu=0.1$ remains consistently stable around an ECS value of approximately $1.0$ until $60\%$ of nodes are removed. Subsequently, between $60\%$ and $85\%$ of nodes removed, the curve gradually and steadily decreases from $1.0$ to $0.84$ (i.e., $16\%$ decrease). In contrast, depicted in \Cref{2sw}(a) using $32$-dimensional embedding, the curve representing $\mu=0.1$ remains stable around $1.0$ only until $25\%$ of node are removed. Then, between $25\%$ and $85\%$ of nodes removed, the curve decreases more sharply from $1.0$ to $0.53$ (i.e., $47\%$ decrease).

\subsubsection{Locally Linear Embedding (LLE)}
\begin{figure}[htbp] \centering
\includegraphics[width=8.3cm]{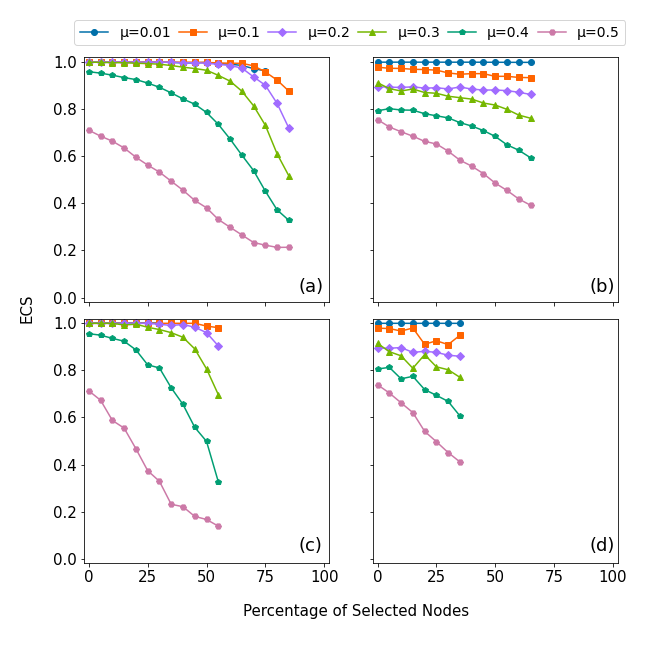}
\caption{Mean ECS using LLE method over the percentage of removed nodes on LFR benchmark graphs with $16$-dimensional embeddings. (a) and (b) apply random node removal, whereas (c) and (d) use targeted removal. Networks in (a) and (c) have $1,000$ nodes, and those in (b) and (d) have $10,000$ nodes.}
\label{3sk}
\end{figure}

\begin{figure}[htbp] \centering
\includegraphics[width=8.3cm]{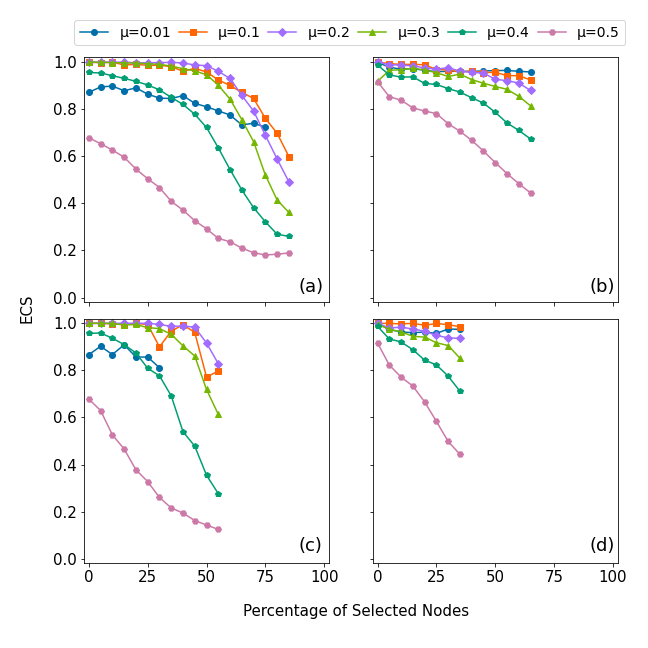}
\caption{Mean ECS using LLE method over the percentage of removed nodes on LFR benchmark graphs with $32$-dimensional embeddings. (a) and (b) apply random node removal, whereas (c) and (d) use targeted removal. Networks in (a) and (c) have $1,000$ nodes, and those in (b) and (d) have $10,000$ nodes.}
\label{3sw}
\end{figure}

Our experimental results on the impact of node removal, generated through the LLE embedding method with dimensions $16$ and $32$, are reported in \Cref{3sk} and \Cref{3sw}, respectively. 

The insights drawn from \Cref{3sk,3sw} for LLE align closely with the insights derived for LE. In fact, the curves in \Cref{3sk} are slightly higher in position than those in \Cref{2sk}, indicating that LLE outperforms LE.
For example, the data points on the green curve (corresponding to $\mu=0.4$) from LLE in \Cref{3sk}(a) have higher ECS scores 
over LE in \Cref{2sk}(a) for a given percentage of removed nodes.

\subsubsection{Higher-Order Preserving Embedding (HOPE)}
\begin{figure}[htbp] \centering
\includegraphics[width=8.3cm]{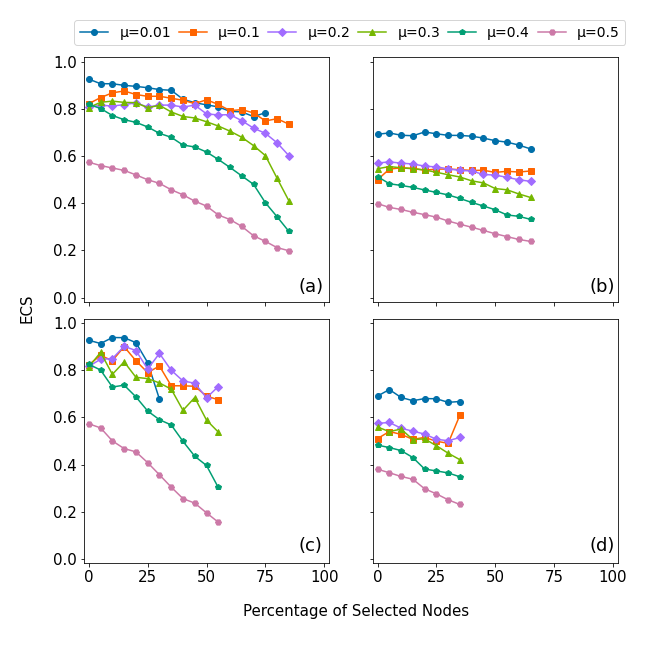}
\caption{Mean ECS using HOPE method over the percentage of removed nodes on LFR benchmark graphs with $16$-dimensional embeddings. While (a) and (b) use random node removal, (c) and (d) rely on targeted removal. Networks in (a) and (c) contain $1,000$ nodes, while those in (b) and (d) consist of $10,000$ nodes.}
\label{1sk}
\end{figure}

\begin{figure}[htbp] \centering
\includegraphics[width=8.3cm]{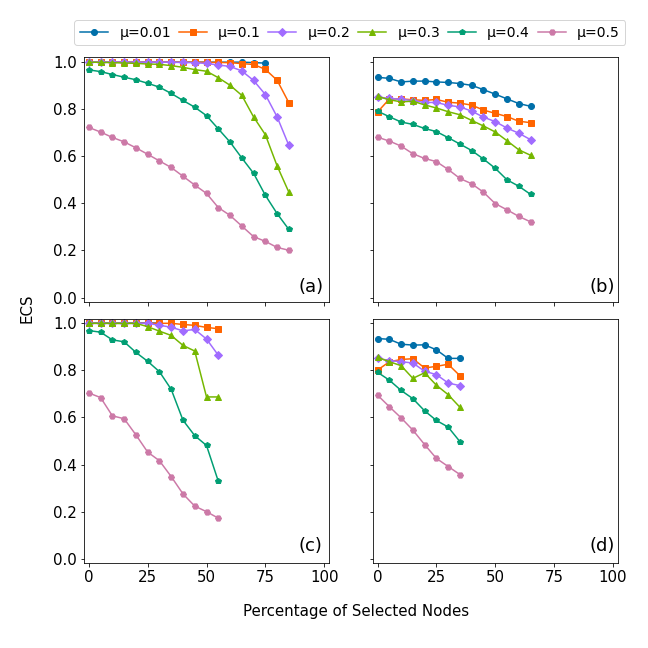}
\caption{Mean ECS using HOPE method over the percentage of removed nodes on LFR benchmark graphs with $32$-dimensional embeddings. While (a) and (b) use random node removal, (c) and (d) rely on targeted removal. Networks in (a) and (c) contain $1,000$ nodes, while those in (b) and (d) consist of $10,000$ nodes.}
\label{1sw}
\end{figure}

Our experimental results on the impact of node removal, generated through the HOPE embedding method with dimensions $16$ and $32$, are reported in \Cref{1sk} and \Cref{1sw}, respectively. 

For the HOPE embedding method, our results suggest that using a 32-dimensional embedding (i.e., \Cref{1sw}) brings higher ECS scores than a 16-dimensional one (i.e., \Cref{1sk}). 
For example, the data points on the green curve (corresponding to $\mu=0.4$) in \Cref{1sw}(a) have higher ECS scores than those data points on the green curve  (corresponding to $\mu=0.4$) in \Cref{1sk}(a) at respective percentage of removed nodes.
We can also see that HOPE has higher ECS scores with smaller graphs.

\subsubsection{Modularized Non-Negative Matrix Factorization (M-NMF)}
\begin{figure}[htbp] \centering
\includegraphics[width=8.3cm]{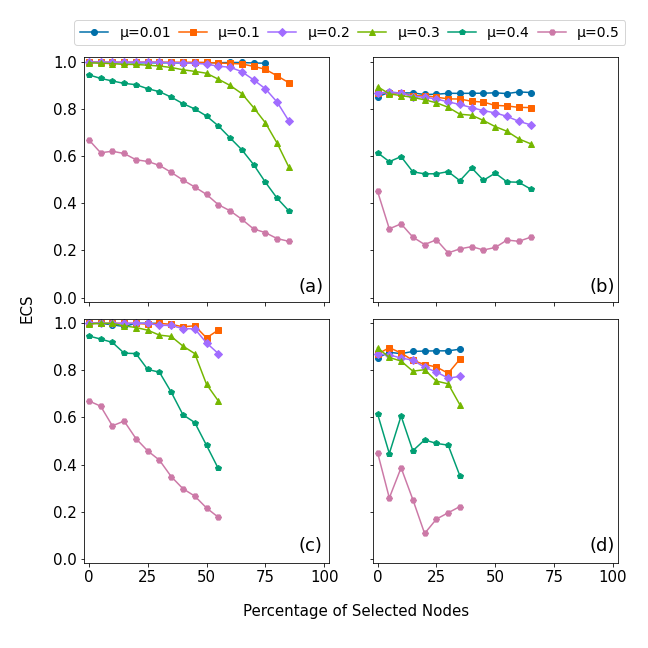}
\caption{Mean ECS using M-NMF method over the percentage of removed nodes on LFR benchmark graphs with $16$-dimensional embeddings. (a) and (b) implement random node removal, while (c) and (d) apply targeted removal; (a) and (c) correspond to networks with $1,000$ nodes, and (b) and (d) to networks with $10,000$ nodes.}
\label{5sk}
\end{figure}

\begin{figure}[htbp] \centering
\includegraphics[width=8.3cm]{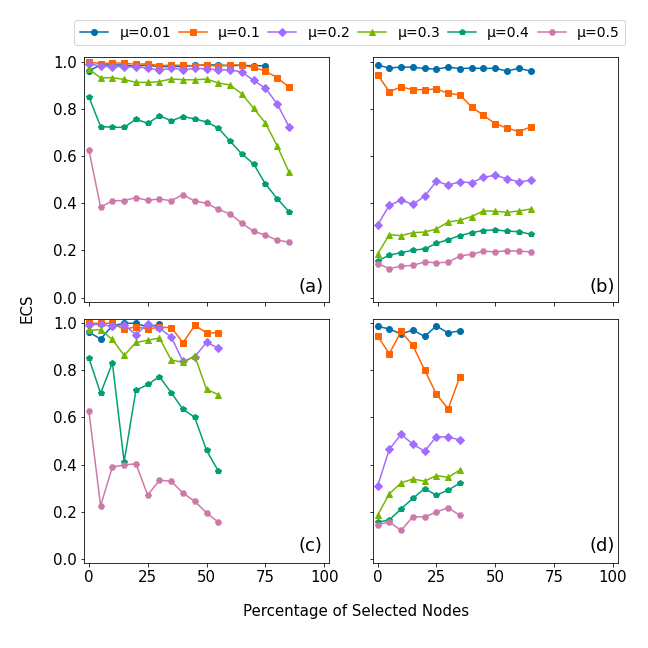}
\caption{Mean ECS using M-NMF method over the percentage of removed nodes on LFR benchmark graphs with $32$-dimensional embeddings. (a) and (b) implement random node removal, while (c) and (d) apply targeted removal; (a) and (c) correspond to networks with $1,000$ nodes, and (b) and (d) to networks with $10,000$ nodes.}
\label{5sw}
\end{figure}

Our experimental results on the impact of node removal, generated through the M-NMF embedding method with dimensions $16$ and $32$, are reported in \Cref{5sk} and \Cref{5sw}, respectively. 

From \Cref{5sk,5sw}, we can see that M-NMF produces higher ECS scores with smaller graphs.

\begin{table*}[htbp]
\caption{\label{tab:table3}%
Choice of more robust graph embedding methods in the family of matrix-factorization.}
\centering
\begin{tabular}{@{}p{0.16\linewidth}|p{0.4\linewidth}|p{0.4\linewidth}@{}}
\toprule
& \textbf{Weaker Community Structure (larger $\mu$)} & \textbf{Stronger Community Structure (smaller $\mu$)} \\
\midrule
\textbf{Smaller Network (smaller $N$)} & LLE with a low embedding dimension & LLE with a low embedding dimension, or HOPE with a high embedding dimension \\
\midrule
\textbf{Larger Network (larger $N$)} & LE with a higher embedding dimension & LE or LLE with a lower embedding dimension \\
\bottomrule
\end{tabular}
\end{table*}

\subsubsection{Summary of Matrix Factorization Methods}
We derive the following insights for matrix factorization methods.
In a nutshell, we observe that ECS scores generally decrease as we progressively remove more nodes and delete them and their adjacent edges and the networks become sparser. This aligns with our intuition that the community structure is more significantly disrupted when a greater number of nodes are removed.
From the figures, it is evident that, in general, for smaller values of $\mu$, the ECS similarity score tends to remain higher for varying degrees of perturbations. 
This suggests that the graph embedding methods tend to produce higher ECS scores when the original LFR benchmark graph exhibits a stronger partition. 
Recall that $\mu$ signifies the fraction of nodes sharing edges across different communities. Thus, a lower $\mu$ value implies a higher ratio of internal to external edges, usually resulting in a more pronounced partition in the LFR benchmark graph.
Notably, when utilizing targeted node removal, the ECS exhibits a more pronounced and rapid decline compared to random removal. 
For example, for LFR network with $1,000$ nodes and $\mu=0.5$, when perturbing the network at $50\%$ and using LE with $16$-dimensional embedding,
we notice that targeted node removal decreases ECS from $0.67$ to $0.14$ (a decrease of $79.1\%$, see \Cref{2sk}(c)), but ECS decreases from $0.67$ to $0.35$ for random node removal (a decrease of $47.8\%$, see \Cref{2sk}(a)).
This observation is comprehensible, as targeted node removal based on betweenness centrality (as a model of adversarial network attack) tends to dismantle the network's community structure more swiftly.

For smaller real-world graphs containing a few thousand vertices, to obtain more robust partitions, the LLE method is suggested with a modest embedding dimension, typically on the order of tens, such as 16 dimensions; Alternatively, when the community partition is stronger, the HOPE embedding method with a higher embedding dimension produces comparably robust partitions.
In the case of larger real-world graphs with a weaker community partition, to obtain more robust community partitions, it is suggested to use the LE embedding method with a moderately higher embedding dimension, such as $32$ 
dimensions as utilized in our experiments (see \Cref{2sw}). 
When dealing with larger real-world graphs characterized by a stronger community structure, the LLE or LE embedding method with a lower embedding dimension produces more robust community partitions. 
In summary, LLE is the most robust method among matrix factorization methods. \Cref{tab:table3} summarizes our findings for matrix factorization methods.

\subsection{Random Walk Methods}
\subsubsection{DeepWalk}
\begin{figure}[htbp]\centering
\includegraphics[width=8.3cm]{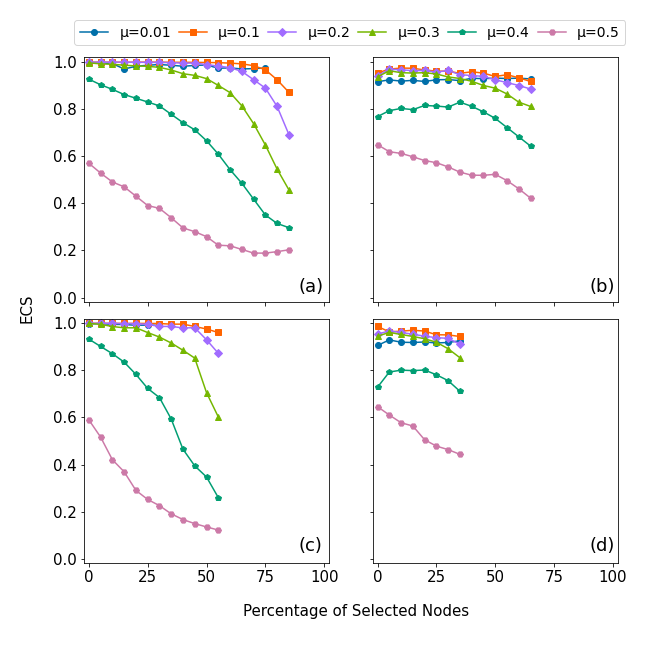}
\caption{Mean ECS using DeepWalk method over the percentage of removed nodes on LFR benchmark graphs with $16$-dimensional embeddings.Random removal is employed in (a) and (b), while targeted removal is used in (c) and (d). (a) and (c) represent networks with $1,000$ nodes, and (b) and (d) represent those with $10,000$ nodes.}
\label{4sk}
\end{figure}

\begin{figure}[htbp]\centering
\includegraphics[width=8.3cm]{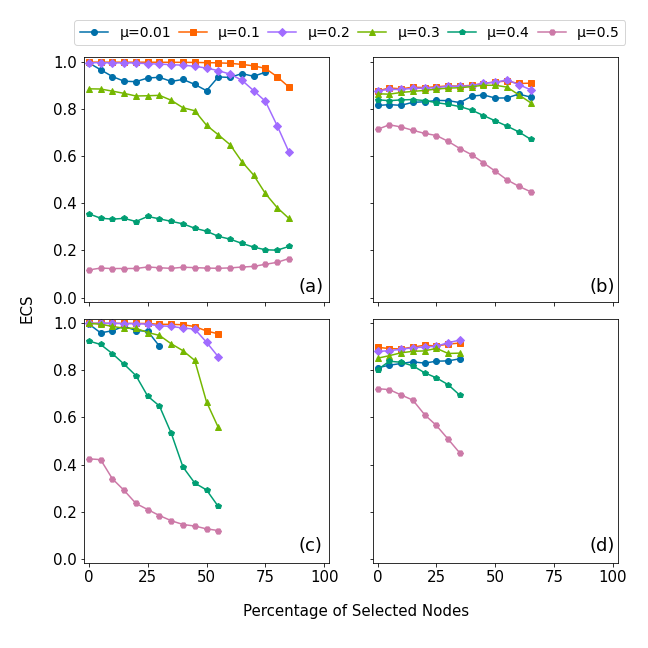}
\caption{Mean ECS using DeepWalk method over the percentage of removed nodes on LFR benchmark graphs with $32$-dimensional embeddings. Random removal is employed in (a) and (b), while targeted removal is used in (c) and (d). (a) and (c) represent networks with $1,000$ nodes, and (b) and (d) represent those with $10,000$ nodes.}
\label{4sw}
\end{figure}

Our experimental results on the impact of node removal, generated through the DeepWalk embedding method with dimension $16$, are reported in \Cref{4sk}. 
Our experimental results using DeepWalk embedding method with dimension $32$ are collected in \Cref{4sw}.

From \Cref{4sk,4sw} for DeepWalk, we can observe that networks with stronger initial partitions (i.e., lower $\mu$ values) experience less perturbation impact and targeted node removal has a significant influence on clusters. Moreover, from the right column of \Cref{4sw}, we see that some curves exhibit unexpected increase as we progressively removed more nodes.

\subsubsection{Large-Scale Information Network Embedding (LINE)}
\begin{figure}[htbp] \centering
\includegraphics[width=8.3cm]{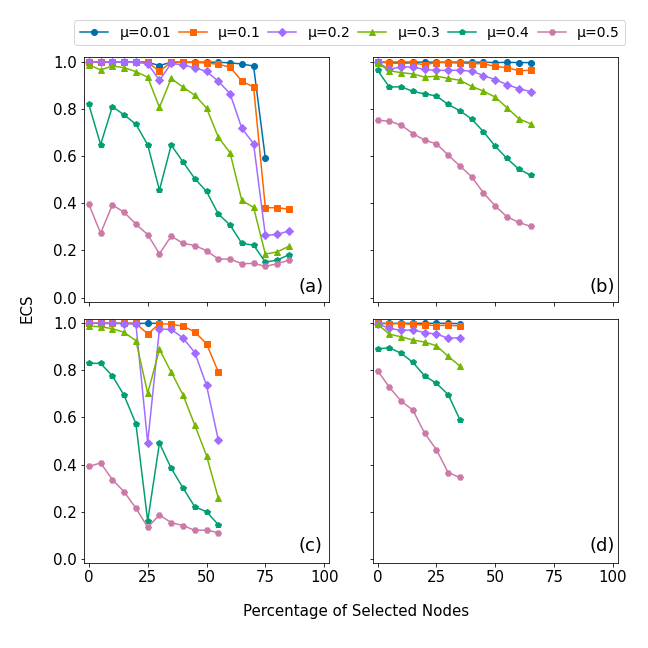}
\caption{Mean ECS using LINE method over the percentage of removed nodes on LFR benchmark graphs with $16$-dimensional embeddings. (a) and (b) follow random node removal, whereas (c) and (d) adopt targeted node removal. Networks with $1,000$ nodes correspond to (a) and (c), and those with $10,000$ nodes correspond to (b) and (d).}
\label{6sk}
\end{figure}

\begin{figure}[htbp] \centering
\includegraphics[width=8.3cm]{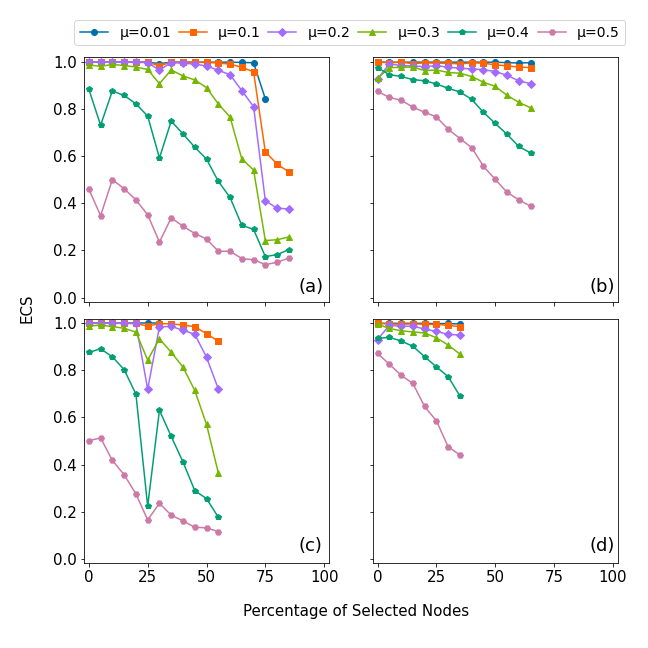}
\caption{Mean ECS using LINE method over the percentage of removed nodes on LFR benchmark graphs with $32$-dimensional embeddings. (a) and (b) follow random node removal, whereas (c) and (d) adopt targeted node removal. Networks with $1,000$ nodes correspond to (a) and (c), and those with $10,000$ nodes correspond to (b) and (d).}
\label{6sw}
\end{figure}

Our experimental results on the impact of node removal, generated through the LINE embedding method with dimensions $16$ and $32$, are reported in \Cref{6sk} and \Cref{6sw}, respectively. 

\begin{figure} \centering
\includegraphics[width=8.3cm]{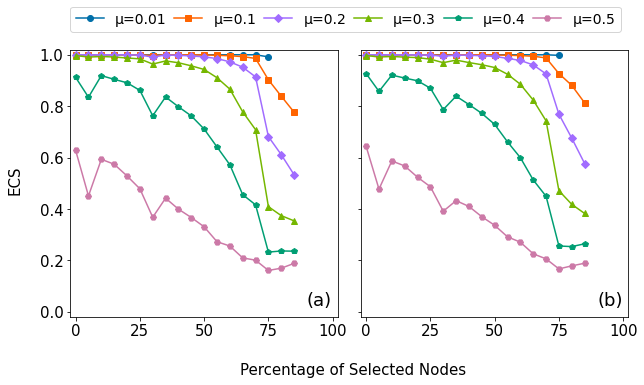}
\caption{Mean ECS using LINE method over the percentage of removed nodes uniformly at random on LFR benchmark graphs with $1,000$ nodes. The first plot displays results using $128$-dimensional embeddings; the second plot displays results using $256$-dimensional embeddings.}
\label{line_high}
\end{figure}

From \Cref{6sk,6sw}. the LINE embedding method produces higher ECS scores when applied to larger networks. Our analysis of networks with 1,000 nodes in the first column of \Cref{6sk} reveals that curves exhibit a steep decline followed by a subsequent increase, but we notice that increasing the embedding dimensions in LINE is expected to mitigate the sharpness of this decline and increase, rendering the pattern less pronounced, as indicated in \Cref{line_high}.
One possible explanation for this abrupt decline is that low embedding dimensions may compress multiple communities into overlapping regions of the vector space. As perturbations remove certain nodes, this compressed structure can initially deteriorate sharply, leading to a sudden drop in ECS; however, further removal of inter-community connections may partially reduce the overlap, resulting in complex non-monotonic behavior in the recovered community structure. As illustrated in Fig.~14, this effect is substantially mitigated when higher embedding dimensions are used, indicating that the phenomenon is related to limited representational capacity.
Based on the aforementioned observation, LINE becomes a viable option for network clustering when dealing with relatively larger networks and the computational resources, particularly GPUs, are available.

\subsubsection{node2vec}
\begin{figure}[htbp] \centering
\includegraphics[width=8.3cm]{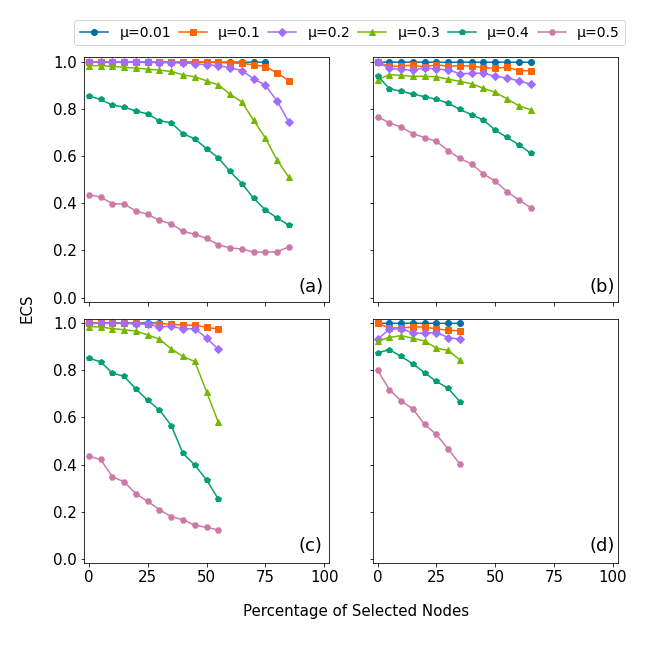}
\caption{Mean ECS using node2vec method over the percentage of removed nodes on LFR benchmark graphs with $16$-dimensional embeddings. Random node removal is applied in (a) and (b), while (c) and (d) use targeted removal. (a) and (c) deal with $1,000$-node networks, and (b) and (d) with $10,000$-node networks.}
\label{7sk}
\end{figure}

\begin{figure}[htbp] \centering
\includegraphics[width=8.3cm]{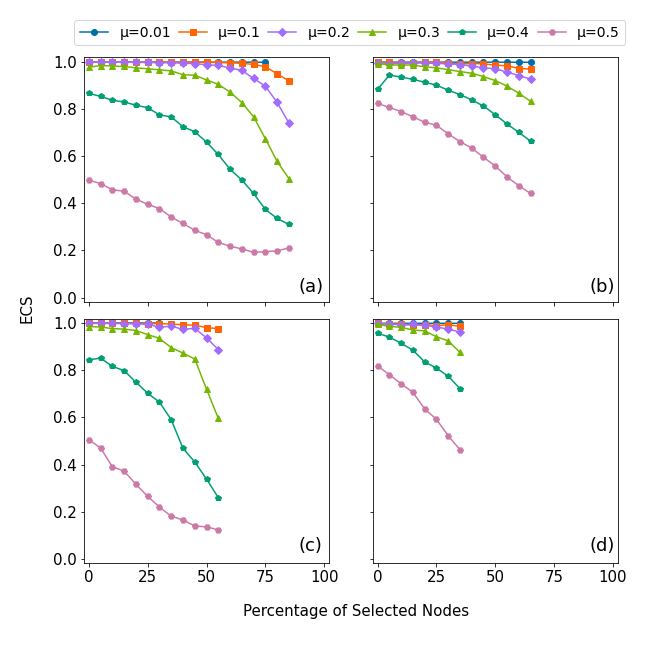}
\caption{Mean ECS using node2vec method over the percentage of removed nodes on LFR benchmark graphs with $32$-dimensional embeddings. Random node removal is applied in (a) and (b), while (c) and (d) use targeted removal. (a) and (c) deal with $1,000$-node networks, and (b) and (d) with $10,000$-node networks.}
\label{7sw}
\end{figure}

Our experimental results on the impact of node removal, generated through the node2vec embedding method with dimensions $16$ and $32$, are reported in \Cref{7sk} and \Cref{7sw}, respectively.

In the case of node2vec, we consistently observe a decrease in ECS as we iteratively remove additional nodes, which aligns with our intuitive expectations. 

\subsubsection{Summary of Random Walk Methods}
Similar to matrix factorization methods, the curve tends to exhibit higher values for smaller mixing parameters ($\mu$). Moreover, when using targeted node removal, the ECS shows a more pronounced and rapid decline compared to random node removal, as evident from the comparison between \Cref{7sk}(a) and \Cref{7sk}(c).

From our discussion of each method, we observe that DeepWalk may exhibit an unexpected increase while LINE experiences unexpected oscillations. In contrast, node2vec is free from these issues and consistently achieves the highest ECS similarity scores. See the Appendix \ref{apdx_n2v} for a brief explanation on the advantage of Node2vec embedding in network community detection.
   
When considering the application of random walk embeddings for community detection, node2vec consistently proves to be a more robust choice over DeepWalk and LINE, across networks of varying sizes, the strength of the initial community partition, and perturbation type.

\subsection{Real-World Networks}\label{sec:real}
In this section, we present the results of our experiments conducted on two real-world networks with labeled communities, i.e., the email-EU-core network and the AS network. 
The email-EU-core network is constructed using email data sourced from a prominent European research institution, comprising $986$ nodes and $16,687$ edges.
{For this network, we use the departmental affiliations of its members as the underlying community structure \cite{Jure-2007}.}
The AS network, derived from the AS Internet topology data collected in June 2009 by the Archipelago active measurement infrastructure, represents the inter-domain Internet topology with Autonomous Systems (ASes) as nodes and AS peerings as links, forming an AS-level topology graph. This dataset comprises $23,748$ nodes and $58,414$ edges.
For this network, we use the densely connected AS groups as the underlying community structure \cite{Boguna-2010}.
To provide additional structural context for the empirical datasets used in our experiments, Table~\ref{table:network properties} reports the principal network statistics of the largest connected component of each network. These measures summarize key structural characteristics including density, average degree, characteristic path length, assortativity, clustering structure, and community organization.
Following the Local Community Paradigm introduced by~\cite{Cannistraci:2013:Link}, we also report the LCP-correlation, a topological measure that quantifies the strength of local community organization in complex networks. The corresponding values for the two empirical networks are summarized in \Cref{table:network properties}.
\begin{table}[!ht]
\centering
\caption{Structural properties of the real-world networks used in the experiments. Statistics are computed on the largest connected component of each network.}
\begin{tabular}{lcc}
\hline
Metric & Email-Eu-core & AS Network \\
\hline
Number of nodes & 986 & 23,748 \\
Number of edges & 25,552 & 58,414 \\
Density & 0.03436 & 0.000207 \\
Average degree & 33.85 & 4.919 \\
Characteristic path length & 2.5869 & 3.5215 \\
Assortativity & -0.0119 & -0.19197 \\
Average clustering coefficient & 0.40705 & 0.36040 \\
LCP-correlation & 0.9631 & 0.9446 \\
Modularity (ground-truth communities) & 0.31304 & 0.17081 \\
Number of communities & 42 & 176 \\
Average community size & 23.476 & 134.932 \\
\hline
\end{tabular}
\label{table:network properties}
\end{table} 

We select them because they closely resemble power-law degree and community size distributions, as the synthetic networks counterpart (i.e., LFR benchmark graphs).
To confirm the alignment of these real-world networks with the structural heterogeneity of LFR networks, we conducted statistical tests on the degree and community size distributions using the Python package \texttt{powerlaw}, following the methodology proposed by Clauset et al.~\cite{Clauset-2009}.
For the email-EU-core network, the estimated scaling exponents are $\alpha = 2.89$ for the degree distribution and $\alpha = 2.76$ for the community size distribution. However, likelihood-ratio tests comparing the power-law model with an exponential alternative do not indicate statistically significant preference for the power-law model ($p > 0.1$), suggesting heterogeneous tail behavior but weaker statistical support for strict power-law scaling.
In contrast, the AS network exhibits strong heavy-tailed characteristics. The degree distribution yields $\alpha = 2.17$ with likelihood-ratio statistic $R = 7.97$ and $p = 1.57 \times 10^{-15}$ when compared with the exponential model, while the community size distribution yields $\alpha = 1.83$, $R = 2.78$, and $p = 0.005$. These results indicate statistically significant support for power-law behavior in both distributions, consistent with the scale-free structure commonly observed in Internet topology networks.
These structural characteristics indicate that the empirical networks exhibit heterogeneous connectivity patterns similar to those modeled in the LFR benchmark graphs, making them suitable testbeds for evaluating the robustness of graph embedding methods under structural perturbations.

As discussed in Section \ref{sec:pert}, in experiments involving LFR networks, we ensure that the removal of each set of nodes does not disconnect the remaining network---a prerequisite for graph embedding methods. However, such a stringent requirement is often impractical for real-world networks. For instance, randomly removing $5\%$ of nodes in the email-EU-core network would typically leave the remaining portion of the network disconnected.
To address this issue in our experiments with real-world networks, we adopt an alternative approach. Specifically, we randomly introduce edges between connected components to restore connectivity. Subsequently, we proceed with the embedding of the reconnected network.

\begin{figure}[htbp] \centering
\includegraphics[width=8.3cm]{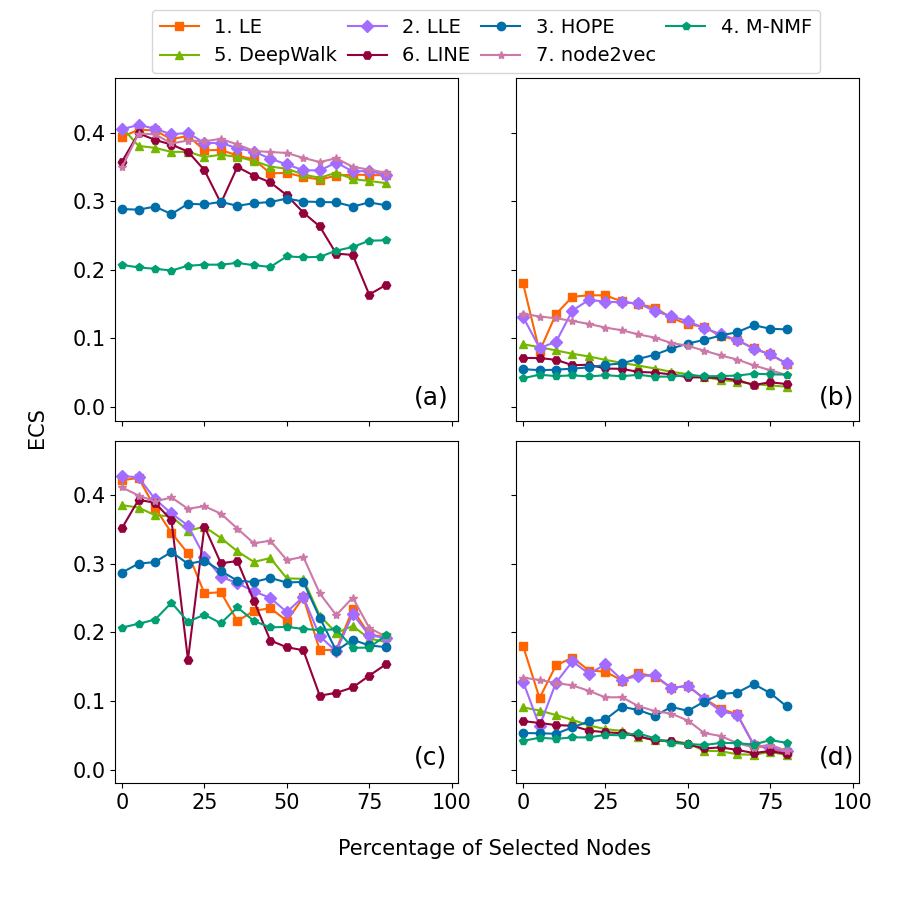}
\caption{Mean ECS for various methods over the percentage of removed nodes on the two real-world networks with $16$-dimensional embeddings. (a) and (b) use random node removal while (c) and (d) use targeted node removal. (a) and (c) corresponds to the email-EU-core network while (b) and (d) corresponds to the AS network.}
\label{986s}
\end{figure}

\begin{figure}[htbp] \centering
\includegraphics[width=8.3cm]{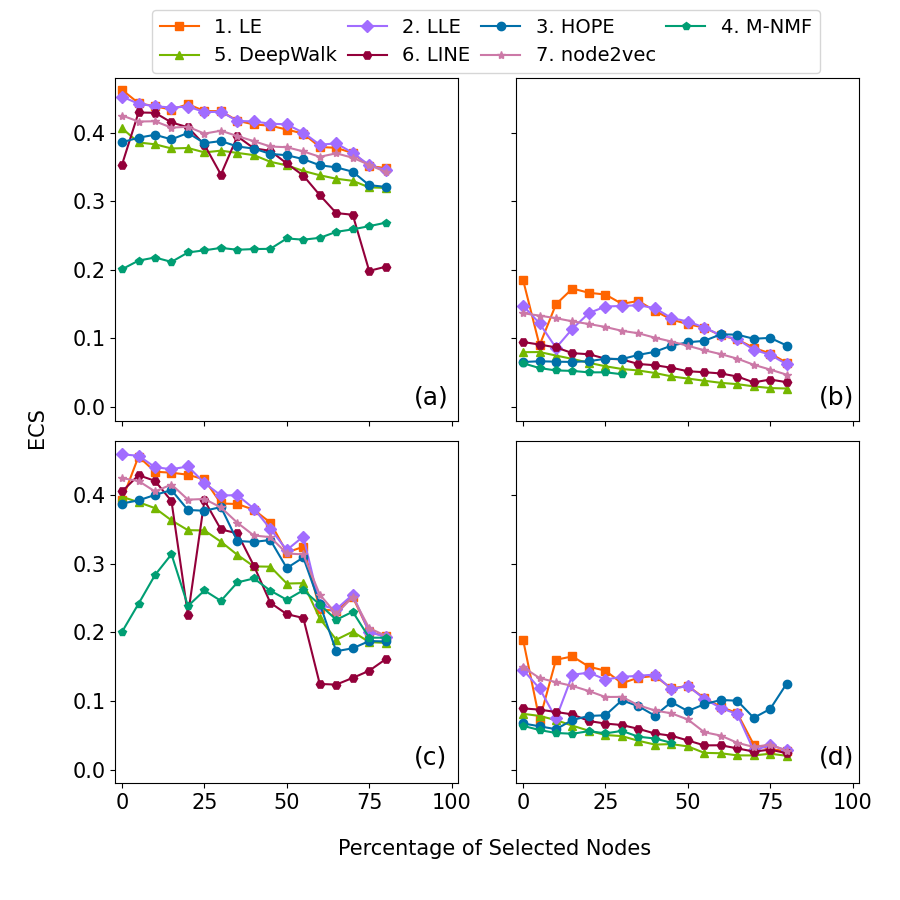}
\caption{Mean ECS for various methods over the percentage of removed nodes on the two real-world networks with $32$-dimensional embeddings. (a) and (b) use random node removal while (c) and (d) use targeted node removal. (a) and (c) corresponds to the email-EU-core network while (b) and (d) corresponds to the AS network.}
\label{23748s}
\end{figure}

We report experimental results using different methods with $16$-dimensional embedding on these two real-world networks in \Cref{986s}. 
The experimental results with $32$-dimensional embedding on the two real-world networks are depicted in \Cref{23748s}.

Comparing the two rows of \Cref{986s}, we can see that when employing targeted node removal, the ECS experiences a more significant decrease than the case of random node removal.
For example, for the email-EU-core network, when perturbing it at $50\%$ and using node2vec with $16$-dimensional embedding,
we notice that targeted node removal decreases ECS from $0.41$ to $0.31$ (a decrease of $24.3\%$, see \Cref{986s}(c)), but ECS is only decreased by around $0.02$ by random node removal (\Cref{986s}(a)).
Furthermore, LLE and node2vec proved to be the superior choices among all seven shallow embedding methods. In general, curves corresponding to these two methods in \Cref{986s} are higher in position than other curves.

As the number of removed nodes increases, the ECS sometimes rises, for example, in \Cref{986s}(a) and \Cref{23748s}(a) for the M-NMF method. To get more insights and mitigate this pathological increasing behavior, we have performed additional experiment and analyzed the ECS score of M-NMF with 128-dimensional embeddings. 
We notice that for clustering bigger networks where the expected number of communities is naturally higher, M-NMF required the use of a higher embedding dimension to effectively separate those communities. 
See the Appendix \ref{apdx} for more details.

{For the two real-world networks we analyzed, the ``underlying community structure'' is assigned based on the functional roles and real-world context of each network. While this assignment is somewhat artificial, we believe it is meaningful and provides a reasonable approximation. By leveraging these predefined community structures, we can still gain valuable insights into the robustness of our community detection methods using network embeddings.}

\subsection{Quantitative Comparison of shallow Graph Embedding Methods}
Without loss of generality, consider a network with $10,000$ nodes, embedding dimension $32$, random node removal, and mixing parameter $\mu=0.1$. The experiments described in our manuscript generate an array of ECS similarity scores for each embedding method, as we progressively remove a larger percentage of nodes. Specifically, for LLE embedding method, this array of ECS similarity scores are presented as the orange curve in \Cref{3sw}(b) in the manuscript. 
For each method, we record the starting ECS value $y_0$, defined as the similarity score when no nodes are removed. To quantify the stability of each embedding as nodes are progressively removed, we define the mean decay rate $\Delta y$ as follows. Let the ECS similarity scores corresponding to node removal levels of $0\%, 5\%, 10\%, \dots, 65\%$ be denoted as $y_0, y_1, \dots, y_{13}$. Then, the decay rate is computed as the average absolute difference between consecutive ECS values:
\begin{displaymath}
    \Delta y = \frac{1}{13} \sum_{i=1}^{13} |y_i - y_{i-1}|.
\end{displaymath}
This quantity captures the average rate of decline in similarity as more nodes are removed. A smaller $\Delta y$ indicates that the embedding method better preserves the community structure under perturbation, i.e., it is more robust.
The mean decay rate of ECS quantifies how rapidly the recovered community structure deteriorates as perturbations increase, providing a direct measure of robustness to progressive structural degradation.
Naturally, a more robust embedding method is characterized by having a larger $y_0$ (indicating better initial similarity) and a smaller $\Delta y$ (indicating better capacity to ``maintain'' community structure as more nodes are removed).

To provide a comprehensive evaluation, we explore networks with varying node counts ($1,000$ or $10,000$) and different mixing parameters ($\mu = 0.01, 0.1, 0.2, 0.3, 0.4, 0.5$), and we also conducted experiment with embedding dimensions of $16$ or $32$, for both random and targeted node removal. Given the variety of parameters, we compute for each embedding method the mean values over these configurations. We summarize the overall robustness of community detection across these seven shallow graph embedding methods in \Cref{tab3}.

\begin{figure}[H]
\centering
\includegraphics[width=0.45\textwidth]{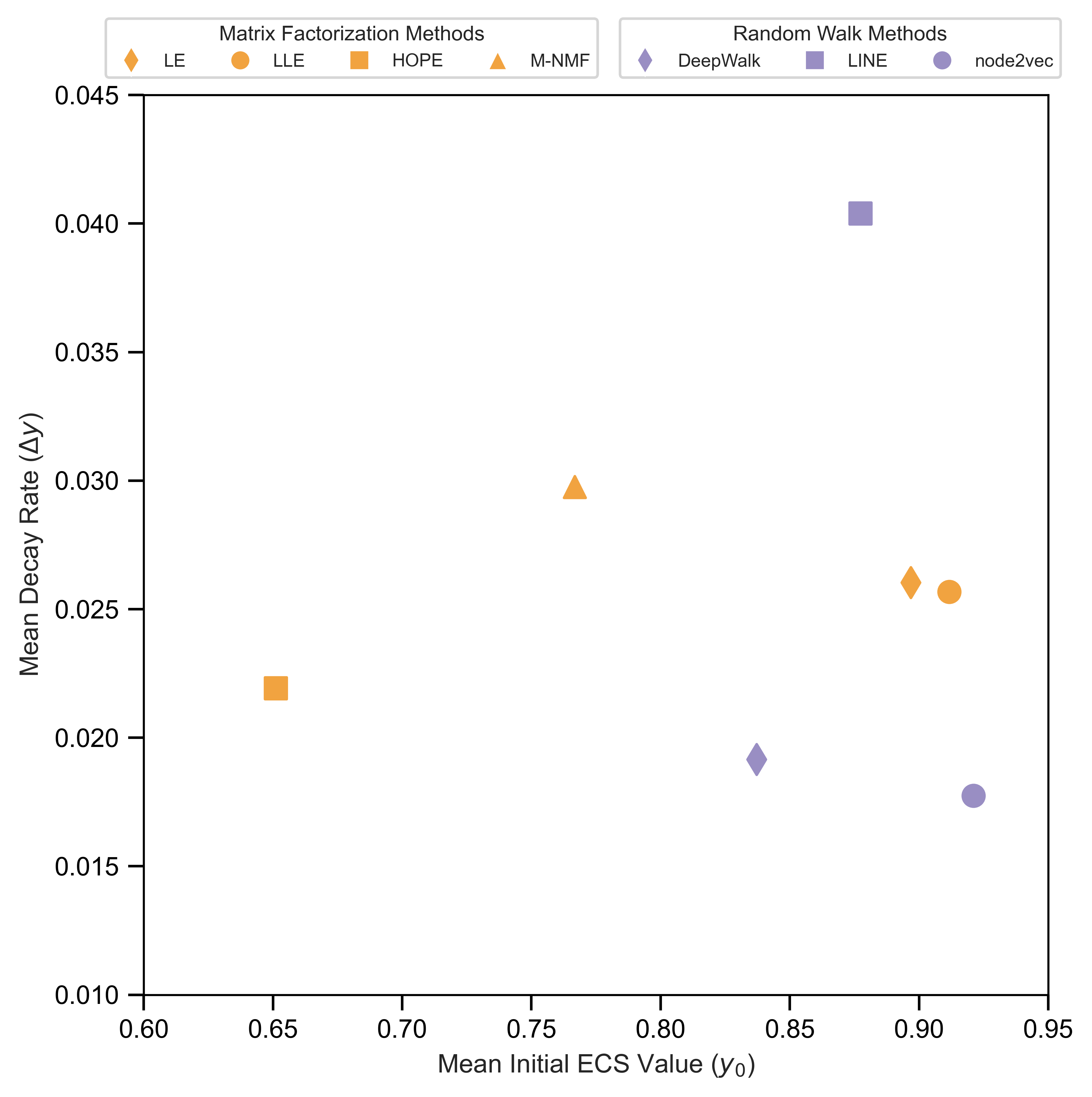}
\caption{Visualization of overall robustness metrics ($y_0$ and $\Delta y$) for different shallow embedding methods, averaged across all configurations including {both random and targeted node removal}, various network sizes, mixing parameters, and embedding dimensions.}
\label{fig3}
\end{figure}

\begin{table*}[htbp]
\caption{\label{tab3}%
Mean robustness metrics for different shallow embedding methods, averaged over multiple experimental settings including varying network sizes ($1{,}000$ and $10{,}000$ nodes), mixing parameters ($\mu = 0.01$ to $0.5$), embedding dimensions ($16$ and $32$), and both random and targeted node removal strategies.
Bold values indicate the robustness metrics of the method identified as the most robust within each embedding family, based on the joint consideration of $y_0$ (higher is better) and $\Delta y$ (lower is better).
}
\centering
\begin{tabular}{lccccccc}
\toprule
\textbf{Family} & \multicolumn{4}{c}{\textbf{Matrix Factorization Methods}} & \multicolumn{3}{c}{\textbf{Random Walk Methods}} \\
\cmidrule(lr){2-5} \cmidrule(lr){6-8}
\textbf{Methods} & LE & LLE & HOPE & M-NMF & DeepWalk & LINE & node2vec \\
\midrule
$y_0$    & 0.896 & \textbf{0.912} & 0.651 & 0.766 & 0.837 & 0.877 & \textbf{0.921} \\
$\Delta y$ & 0.026 & \textbf{0.025} & 0.022 & 0.029 & 0.019 & 0.040 & \textbf{0.017} \\
\bottomrule
\end{tabular}
\end{table*}
From \Cref{tab3}, we can confirm that LLE and node2vec provide the most robust community detection results, which aligns with our qualitative analysis in the manuscript. 
Note that, in \Cref{tab3}, LLE has larger $\Delta y$ than HOPE; however, since LLE has a much larger $y_0$ than HOPE, we still identify LLE as the best method within the family of matrix-factorization methods.
To make it more visually convenient for the readers, we have visualize \Cref{tab3} as \Cref{fig3}. In \Cref{fig3}, points near the bottom-right corner correspond to more robust embedding methods. Again, LLE and node2vec outperform other embedding methods and show best overall robustness.

We further observe that the robustness conclusions, particularly the consistently strong performance of node2vec and LLE, remain unchanged across the full range of mixing parameter values considered in this study, including $\mu=0.5$, where community boundaries are weakest. Although this study does not explicitly vary the degree and community size exponents of the LFR model, these results suggest that the relative robustness of these methods is maintained even as community structure becomes increasingly difficult to detect. This observation is consistent with the theoretical analysis of~\cite{Kojaku:2023:Community:Detection:Neural:Embedding}, who showed that node2vec successfully identifies communities in stochastic block models, which exhibit substantially more homogeneous degree distributions than LFR benchmark graphs. Together, these findings suggest that the robustness advantage of node2vec is not solely a consequence of the scale-free characteristics of the LFR benchmark.

\subsection{Limitations of Our Study}
Having discussed the insights derived from our results, we acknowledge the following limitations of our work:
\begin{enumerate}
\item Our empirical evaluation is based on LFR benchmark graphs and two real-world networks. While this combination provides both controlled experimentation and validation on empirical data, additional real-world networks from other domains, such as biological, transportation, or social systems, may exhibit robustness behaviors not captured in the datasets considered here. The robustness evaluation framework developed in this study is general and not restricted to these datasets. To facilitate further investigation, we have publicly released all code and data used in this work, enabling researchers to reproduce and extend the proposed analysis to additional network domains.
\item Our study emphasizes computational investigations of community detection robustness rather than analytical research. Consequently, we do not analyze detectability limits, which remain underexplored for LFR benchmark graphs due to their complex generation and structural properties.
\item In our experiments on two real-world networks, the underlying community structure of ground truth of communities was assigned based on real-world context as determined by domain experts. However, such assignments may introduce subjectivity and potential biases.
\item When we study the two real-world networks, we have artificially reconnected disconnected components by introducing 
minimal random edges to ensure a globally consistent embedding space.
This was necessary due to a technical limitation of the graph embedding methods we employed, which tend to perform poorly on disconnected graphs. 
    In fact, the graph embedding methods we used embed each connected component independently for disconnected graphs, leading to \emph{arbitrary relative positioning} between components, which undermines the global consistency needed for downstream tasks such as clustering or community detection. 
    We acknowledge that artificial reconnection would introduce some deviation from the original network structure.
\end{enumerate}

\section{\label{sec:conclusion}Conclusion}
We conducted a systematic set of experiments aimed at evaluating the robustness of shallow graph embedding methods for community detection on networks. Our study encompassed LFR benchmark graphs and two real-world networks, with variations in network sizes and mixing parameters considered for the LFR benchmark graphs.
We studied two different strategies for perturbations: (1) node removal based on random node removal, analogous to random network errors, and (2) node removal based on targeted node removal, modeling deliberate attacks. Both perturbation strategies involve the removal of removed nodes. Our community detection approach utilized seven graph embedding methods. Community similarity was measured through ECS.

Our experimental findings revealed a general decline in community similarity as more nodes were deleted. Notably, targeted node removal led to a more pronounced and rapid decline in community similarity compared to random node removal.
Analysis of clustering similarity scores suggested that LFR networks with lower mixing parameters, indicative of a stronger community structure, exhibit greater robustness to network perturbation. Moreover, diverse graph embedding methods displayed varying degrees of robustness in network community detection. Specifically, LLE 
demonstrated proficiency within the family of matrix-based methods. For smaller real-world networks with a limited number of vertices, the more robust embedding method is LLE with a modest embedding dimension, typically on the order of tens --- such as the 16 dimensions employed in our experiment. 
In the case of larger real-world networks, the LE and LLE embedding methods showed greater robustness to partition. Specifically, LE with a higher embedding dimension is preferable when the community structure is less pronounced, whereas LE or LLE with a lower embedding dimension is more effective when the network exhibits a stronger community structure.
Within the family of random walk methods, our results consistently highlighted node2vec as the superior choice. Notably, node2vec also outperformed all methods within the family of matrix factorization methods.

In contemplating potential avenues for future research, it is noteworthy that our current experiments did not involve parameter optimization in the graph embedding algorithms, primarily due to its inherent computational expense. To attain a more exhaustive and comprehensive insight into the robustness of community detection using shallow graph embedding methods, we may explore the incorporation of parameter optimization in our forthcoming studies. 
Moreover, while this work focuses on shallow graph embeddings, we acknowledge that neural network-based methods have shown significant promise in recent years. GNNs, particularly in clustering tasks, leverage node features and graph topology in an end-to-end learning framework, providing powerful embeddings that can capture complex relationships within graphs. We plan to explore such approaches in future work to compare their robustness with traditional methods.

\section*{Acknowledgments}
The work of Zhi-Feng Wei is supported by the U.S.\  Department of Energy (DOE) Office of Advanced Scientific Computing Research (ASCR) through the ASCR Distinguished Computational Mathematics Postdoctoral Fellowship (Project No.\ 71268). 
Pacific Northwest National Laboratory (PNNL) is a multi-program national laboratory operated for the U.S. Department of Energy (DOE) by Battelle Memorial Institute under Contract No. DE-AC05-76RL01830.

The research of Zhi-Feng Wei was supported in part by an appointment with the National Science Foundation (NSF) Mathematical Sciences Graduate Internship (MSGI) Program. This program is administered by the Oak Ridge Institute for Science and Education (ORISE) through an interagency agreement between the U.S.\ Department of Energy (DOE) and NSF. ORISE is managed for DOE by ORAU. All opinions expressed in this paper are the author's and do not necessarily reflect the policies and views of NSF, ORAU/ORISE, or DOE.
The research of Zhi-Feng Wei was supported in part by an appointment to the Oak Ridge National Laboratory GRO Program, sponsored by the U.S.\ Department of Energy and administered by the Oak Ridge Institute for Science and Education.

This paper has been coauthored by UT-Battelle, LLC under Contract No.\ DE-AC05-00OR22725 with the U.S.\ Department of Energy. The publisher, by accepting the article for publication, acknowledges that the U.S.\ government retains a nonexclusive, paid-up, irrevocable, world-wide license to publish or reproduce the published form of the manuscript, or allow others to do so, for U.S.\ government purposes. The DOE will provide public access to these results in accordance with the DOE Public Access Plan (http://energy.gov/downloads/doe-public-access-plan).
This material is based upon work supported by the U.S.\ Department of Energy, Office of Science, Office of Advanced Scientific Computing Research under Contract No.\ DE-AC05-00OR22725. The funders had no role in study design, data collection and analysis, decision to publish, or preparation of this manuscript.

This research was supported in part by Lilly Endowment, Inc., through its support for the Indiana University Pervasive Technology Institute. 
\newpage

\clearpage
\begin{appendices}
\section{\label{apdx_n2v} Remarks on node2vec method}
Node2vec produces more robust graph community detection results due to its flexible random walk strategy, which employs biased random walks to provide a trade-off between breadth-first search (BFS) and depth-first search (DFS). As mentioned in \cite{Goyal-2018}, node2vec produces higher-quality and more informative embeddings than DeepWalk, enabling node2vec to preserve community structure as well as structural equivalence between nodes.

There has been theoretical evidence supporting the superiority of node2vec over DeepWalk in community detection tasks. As demonstrated in \cite{Kojaku:2023:Community:Detection:Neural:Embedding,NEURIPS2021_ca954182}, node2vec is notably robust to degree heterogeneity due to its ability to learn degree-agnostic embeddings. In contrast, DeepWalk tends to encode node degree as a dominant feature in the embedding space. This sensitivity makes DeepWalk more susceptible to structural noise in networks with heterogeneous degree distributions, such as those generated by the LFR benchmark model.
These theoretical insights are consistent with our empirical results. As shown in \Cref{4sk} and \Cref{7sk}, when the proportion of removed nodes is small---so the perturbed network retains structural properties close to the original LFR benchmark---the ECS similarity scores achieved by DeepWalk are consistently lower than those by node2vec. A similar pattern is observed in \Cref{4sw} and \Cref{7sw}.
In addition, this theoretical insight is reflected in \Cref{tab3}: the $y_0$ values for DeepWalk are smaller than those for node2vec in these two tables.

Furthermore, Cohen et al.~\cite{PhysRevLett.85.4626} studied the robustness of networks with power-law degree distributions under random breakdowns using percolation theory. Specifically, they showed that if a network has a degree distribution with probability mass function $\{P_k\}$, and a fraction $p$ of its nodes break down, the resulting degree distribution has probability mass function
\begin{displaymath}
    P'_k=\sum_{\ell=k}^{\infty}P_{\ell}\binom{\ell}{k}(1-p)^k p^{\ell-k}.
\end{displaymath}
This transformation preserves the heavy-tailed nature of the original distribution. 
Furthermore, applying this calculation in the case of LFR graphs, we can get a stronger assertion: If  $\{P_k\}$ is power-law distributed with exponent $\gamma$, then $\{P'_k\}$ is asymptotically power-law distributed and has the same exponent $\gamma$. 
To show this, we can first construct a random sum $S_N = \sum^{N}_{i=1}B_i$, where $N$ is a power-law random variable with exponent $\gamma$ and $\{B_i\}_{i=1}^\infty$ is a sequence of independent Bernoulli random variable whose probability of taking value $1$ is $1-p$. Then the probability mass function of $S_N$ is exactly $\{P'_k\}$. According to Theorem~3.1 in \cite{robert}, we have
\begin{displaymath}
    \Pleft{S_N>x} \sim \Pleft{N>\tfrac{x}{1-p}},\quad\text{as } x\to \infty.
\end{displaymath}
Because $N$ is a power-law random variable with exponent $\gamma$ , it follows that 
\begin{displaymath}
    \Pleft{S_N>x} \sim \bigl(\tfrac{x}{1-p}\bigr)^{-\gamma},\quad\text{as } x\to \infty.
\end{displaymath}
Now that $\{P'_k\}$ is the probability mass function of $S_N$, we proved our assertion: if  $\{P_k\}$ is power-law distributed with exponent $\gamma$, then $\{P'_k\}$ is asymptotically power-law distributed with the same exponent $\gamma$.
Hence, the type of random perturbation we considered does not eliminate degree heterogeneity. Consequently, DeepWalk remains vulnerable to degree-based noise after such perturbations, while node2vec continues to perform robustly in community detection.

As discussed in \cite{Kojaku:2023:Community:Detection:Neural:Embedding,Qiu-2018}, LINE can be regarded as a special case of node2vec with a window size $w=1$, meaning that it focuses solely on learning dyadic (i.e., direct) relationships between nodes. This limited temporal context restricts LINE's ability to capture higher-order or indirect structural features. Consequently, LINE exhibits lower robustness than node2vec, particularly in noisy or perturbed networks, because it tends to encode stochastic and noisy pairwise interactions. In contrast, node2vec, by aggregating longer biased random walks, can effectively capture indirect relationships between nodes, making it more resilient in downstream tasks such as community detection.

Moreover, as mentioned in \cite{Kojaku:2023:Community:Detection:Neural:Embedding}, node2vec is robust to network sparsity compared to traditional spectral embedding methods. These observations are corroborated in our experiments. In fact, as we remove more nodes, networks are getting sparsified. Therefore, it is understandable that node2vec is more robust in our experiments.

Hence, node2vec is more robust against random and targeted perturbations in community detection, outperforming other methods in stability under perturbations.

\section{\label{apdx} M-NMF method with higher embedding dimensions}
An increase in ECS can be observed in \Cref{5sw,986s,23748s} for the M-NMF method, where 16-dimensional and $32$-dimensional embedding are used. 

\begin{figure}[htbp] \centering
\includegraphics[width=8cm]{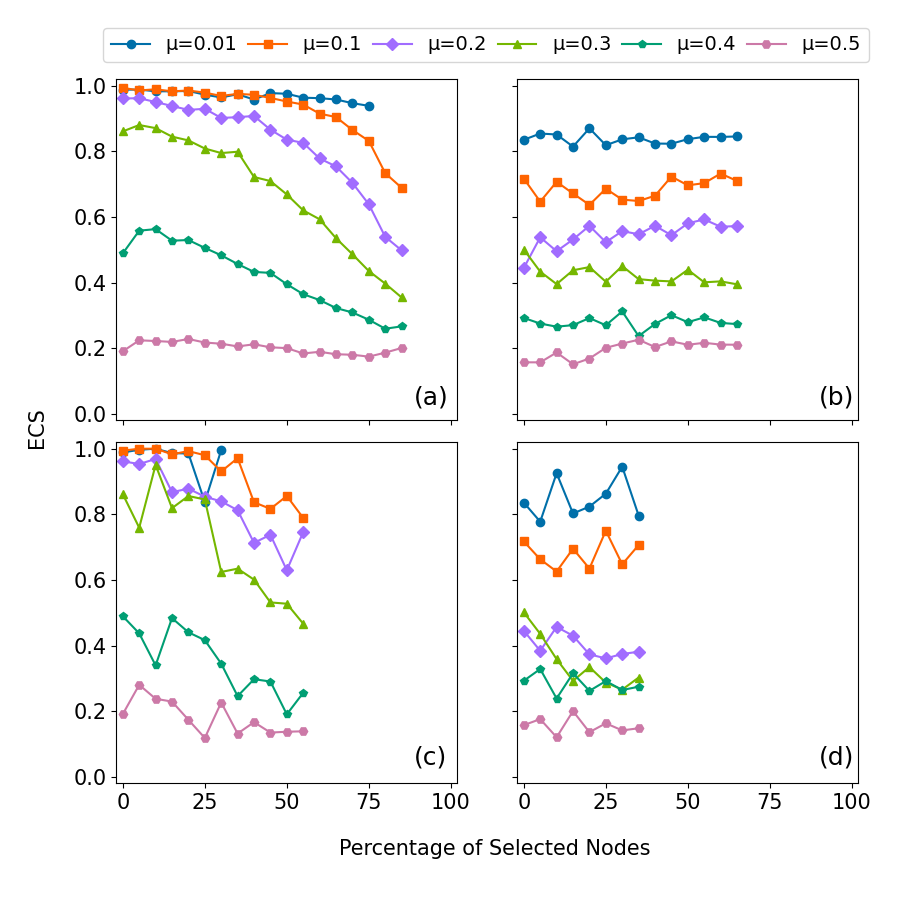}
\caption{Mean ECS using M-NMF method over the percentage of removed nodes on LFR benchmark graphs with $128$-dimensional embeddings. (a) and (b) implement random node removal, while (c) and (d) apply targeted removal; (a) and (c) correspond to networks with $1,000$ nodes, and (b) and (d) to networks with $10,000$ nodes.}
\label{128_5MNMF}
\end{figure}

We see some unexpected increasing behaviors in the M-NMF curves in \Cref{5sw}(b) and \Cref{5sw}(d) for synthetic LFR networks with $10,000$ nodes. Intuitively, these networks are bigger and therefore more complex, which makes it more difficult to detect communities. We thus conducted additional experiments on M-NMF with a $128$-dimensional embedding as shown in \Cref{128_5MNMF}. We observe that the unexpected increasing behaviors are mitigated compared to $32$-dimensional embedding. Specifically, the ECS scores of M-NMF with $128$-dimensional embedding are more stable than those with $32$-dimensional embedding. This indicates that the choice of embedding dimension can significantly and sensitively impact the performance of M-NMF. Due to the substantial computational cost associated with our additional experiments on synthetic networks with $10,000$ nodes, we reduced the number of realizations per percentage of removed nodes to $10$, compared to the $50$ used in the original paper, to obtain results within the high-performance computing cluster's time limit. Standard deviations are still negligible compared to the mean but we do not show them here.

\begin{figure}[htbp] \centering
\includegraphics[width=6.5cm]{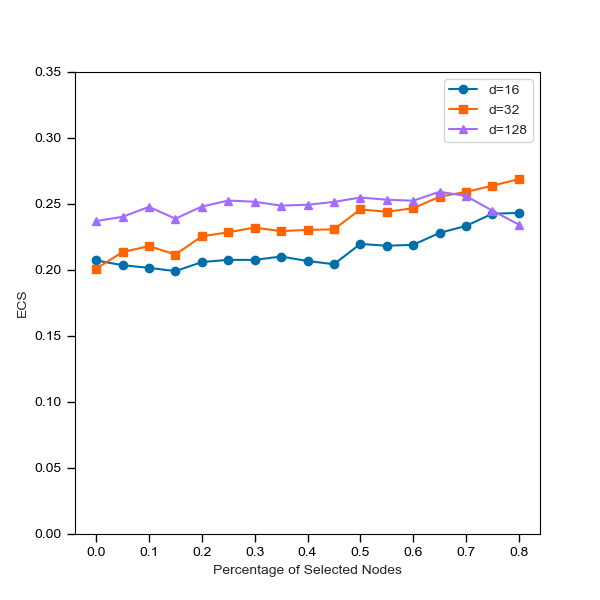}
\caption{Mean ECS using M-NMF method on email-EU-core network with $986$ nodes with different embedding dimensions.}
\label{comparison_results}
\end{figure}

Notably, the M-NMF curves in \Cref{986s}(a) and \Cref{23748s}(a) for email-EU-core networks are generated using the exact same node removal sequence, with the only difference being the embedding dimension. In \Cref{986s}(a), where the embedding dimension is $16$, the M-NMF curve remains almost stable showing slightly unexpected increases. In \Cref{23748s}(a), the increasing of ECS is more noticeable. To get more insights into this behavior, we performed additional experiment and analyzed the ECS scores of M-NMF with $128$-dimensional embedding. The comparison results are presented in \Cref{comparison_results}. We can see that the ECS scores of M-NMF with $128$-dimensional embedding are more ``stable'' as more nodes are removed, where unexpected increasing behaviors are much less pronounced. This again suggests that the choice of embedding dimension can significantly impact the performance of M-NMF, as it has been noticed by \cite{Tandon-2021} and \cite{zhu-1}. Furthermore, we observe a difference in community structure: our synthetic LFR networks with $1,000$ nodes exhibit approximately $10$ communities; our synthetic LFR networks with $10,000$ nodes exhibit approximately $20$ communities. In contrast, the email-EU-core network contains $42$. Consequently, employing a higher embedding dimension, such as $128$, appears reasonable to effectively separate these more numerous communities in the embedding space for email-EU-core network.
    
In summary, we have observed that the choice of embedding dimension can significantly sensitively impact the performance of M-NMF in community detection, which can lead to pathological behaviors in the ECS scores. This reveals that M-NMF is not as robust compared to other methods, especially node2vec, which is more stable and less sensitive to the choice of embedding dimension.

\section{Modularity Evolution Under Node Removal}

In this appendix, we examine how the modularity $Q$ of the network evolves under node removal for the LFR benchmark model. We consider networks with $1{,}000$ nodes and evaluate two attack strategies: random node removal and targeted removal based on betweenness centrality.

Node removal is performed under the constraint that the remaining network must remain connected. Therefore, articulation points that would disconnect the graph are not removed during the attack process.

To investigate the effect of community strength, we consider two mixing parameters in the LFR benchmark: $\mu=0.1$ and $\mu=0.5$. A smaller mixing parameter corresponds to stronger community structure, while larger values indicate weaker separation between communities.

\begin{figure}[htbp]
\centering
\includegraphics[width=0.9\linewidth]{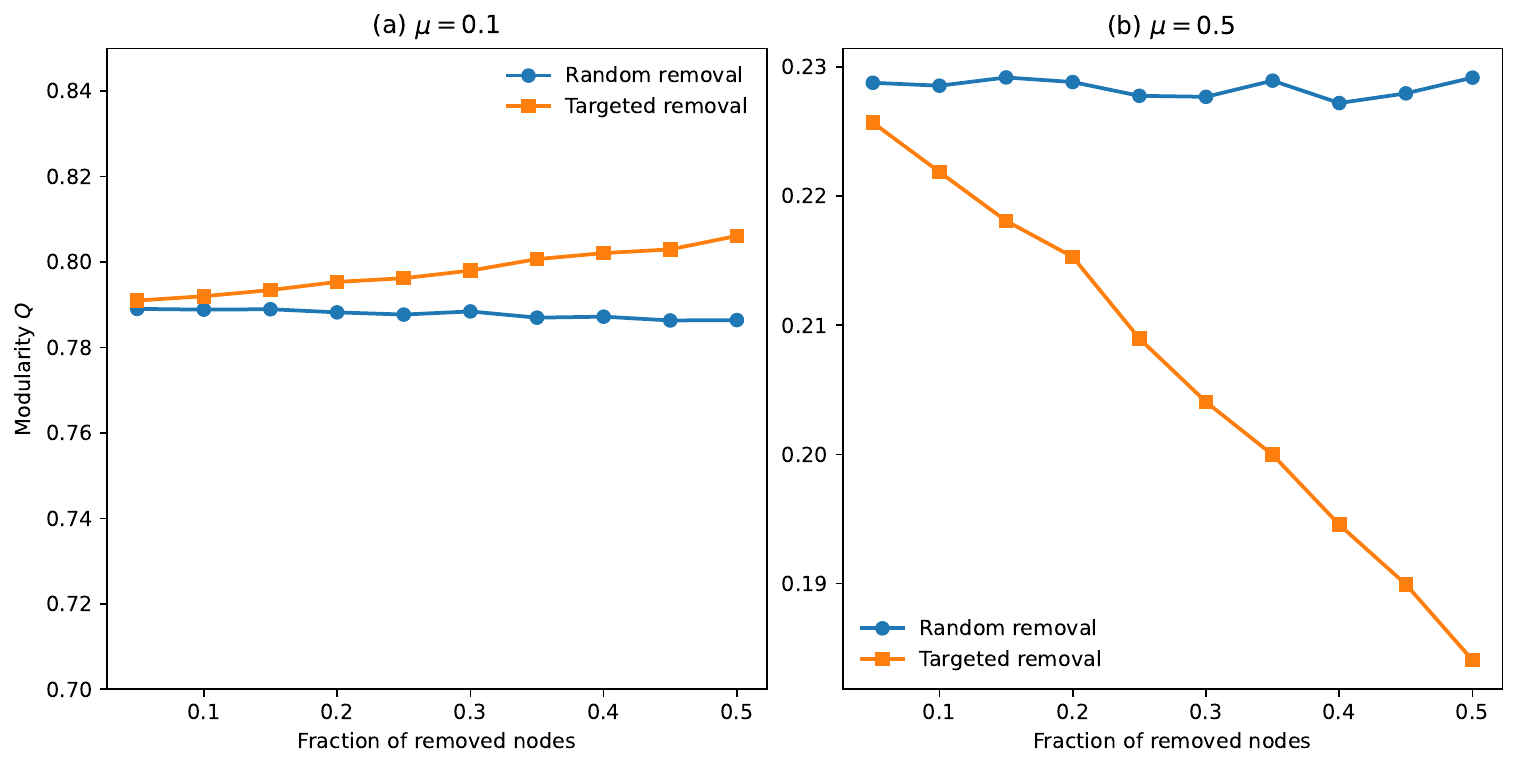}
\caption{
Evolution of modularity $Q$ under node removal for LFR benchmark networks with different mixing parameters. 
(a) $\mu=0.1$, corresponding to strong community structure. 
(b) $\mu=0.5$, corresponding to weaker community structure. 
Each subplot compares random node removal and targeted removal based on betweenness centrality. 
The x-axis represents the fraction of removed nodes and the y-axis shows the modularity of the perturbed network. Each data point represents the average across 50 realizations.
}
\label{fig:lfr_modularity_evolution}
\end{figure}

Figure~\ref{fig:lfr_modularity_evolution} shows how modularity evolves as the fraction of removed nodes increases under the two attack strategies.

When $\mu=0.1$, the network exhibits strong community structure. In this case, both random and targeted node removal lead to relatively stable modularity values. Targeted removal may even slightly increase modularity. This effect occurs because nodes with high betweenness centrality often participate in inter-community connections. Removing these nodes can reduce cross-community links while preserving the internal structure of communities, which may increase the modularity of the remaining graph.

When $\mu=0.5$, the community structure becomes significantly weaker and a larger fraction of edges already connect different communities. Under this regime, high-betweenness nodes are not necessarily bridge nodes between well-separated communities. Removing these nodes may instead disrupt the remaining community structure, leading to a gradual decrease in modularity under targeted node removal.

These observations highlight that the evolution of modularity depends on the strength of the underlying community structure. While modularity reflects the structural separation between communities in the perturbed network, ECS evaluates how accurately the original community assignments can still be recovered after perturbation. Therefore, modularity and ECS provide complementary perspectives on network robustness.

\section{Computational complexity of shallow graph embeddings}
In this appendix, we summarize the asymptotic time complexity (dominant term) of the seven embedding techniques considered in this work. \\

\begin{itemize}

\item \textbf{Laplacian Eigenmaps (LE):}
Laplacian Eigenmaps requires solving a sparse eigenvalue problem associated with the graph Laplacian to obtain the embedding coordinates. 
The original formulation describes the method as involving ``a few local computations and one sparse eigenvalue problem''~\cite{Belkin-2003}. 
Under common implementations, the dominant computational cost arises from this sparse eigensolver step, which scales approximately as $O(|E|\, d^{2})$, where $|E|$ is the number of edges and $d$ is the embedding dimension.

\item \textbf{Locally Linear Embedding (LLE):}
Locally Linear Embedding first computes local reconstruction weights based on nearest neighbors and then obtains the embedding by solving an eigenvalue problem~\cite{Sam-2000}. 
The dominant computational cost arises from the sparse eigenvalue decomposition step, which scales approximately as $O(|E|\, d^{2})$ under common implementations.

\item \textbf{HOPE:}
HOPE (High-Order Proximity preserved Embedding) reduces the high-order proximity preservation problem to a generalized singular value decomposition (GSVD) formulation~\cite{Ou-2016}. 
Using iterative eigensolvers, the dominant computational cost scales approximately as $O(|E|\, d^{2})$, where $|E|$ is the number of edges and $d$ is the embedding dimension.

\item \textbf{M-NMF:}
M-NMF jointly models community structure and node embeddings through a modularized nonnegative matrix factorization framework~\cite{WangX-2017}. 
The algorithm relies on iterative multiplicative update rules for matrix factors and community indicators. 
Under common implementations, the dominant computational cost scales approximately as $O(|V|\,(d+c))$, where $|V|$ is the number of nodes, $d$ is the embedding dimension, and $c$ is the number of communities.

\item \textbf{DeepWalk:}
DeepWalk generates truncated random walks from each node and learns embeddings using the Skip-Gram model with negative sampling~\cite{Perozzi-2014}. 
When walk parameters are treated as constants, the dominant computational cost scales approximately as $O(|V|\, d)$, where $|V|$ is the number of nodes and $d$ is the embedding dimension.

\item \textbf{LINE:}
LINE learns node embeddings by optimizing first- and second-order proximity objectives using edge sampling and negative sampling~\cite{Tang-2015}. 
The optimization process typically performs a number of stochastic gradient updates proportional to the number of edges, leading to an overall computational complexity of approximately $O(|E|\, d)$.

\item \textbf{node2vec:}
node2vec extends DeepWalk by introducing biased second-order random walks to better explore network structure~\cite{Grover-2016}. 
After preprocessing transition probabilities for the random walks, the dominant computational cost arises from random walk generation and Skip-Gram training, which scales approximately as $O(|V|\, d)$ when walk parameters are treated as constants.
\end{itemize}
\end{appendices}

\bibliography{Literature}

@inbook{Bhagat:2011:Node:Classification,
   author = {Bhagat, Smriti and Cormode, Graham and Muthukrishnan, S.},
   title = {Node Classification in Social Networks},
   booktitle = {Social Network Data Analytics},
   editor = {Aggarwal, Charu C.},
   publisher = {Springer US},
   address = {Boston, MA},
   pages = {115-148},
   ISBN = {978-1-4419-8462-3},
   year = {2011},
   type = {Book Section}
}

@misc{Goel:2025:Community:Detection:Robustness:GNN,
  title = {Community detection robustness of graph neural networks},
  author = {Goel, Jaidev and Moriano, Pablo and Kannan, Ramakrishnan and Gel, Yulia R.},
  journal = {Phys. Rev. Res.},
  volume = {8},
  issue = {2},
  pages = {023123},
  numpages = {26},
  year = {2026},
  month = {May},
  publisher = {American Physical Society},
  doi = {10.1103/v6nq-hcm8},
  url = {https://link.aps.org/doi/10.1103/v6nq-hcm8}
}

@ARTICLE{zhu-1,
  author={Zhu, Junyou and Wang, Chunyu and Gao, Chao and Zhang, Fan and Wang, Zhen and Li, Xuelong},
  journal={IEEE Transactions on Network Science and Engineering}, 
  title={Community Detection in Graph: An Embedding Method}, 
  year={2022},
  volume={9},
  number={2},
  pages={689-702},
  doi={10.1109/TNSE.2021.3130321}}

@article{zhu-2,
  title={Unsupervised community detection in attributed networks based on mutual information maximization},
  author={Zhu, Junyou and Li, Xianghua and Gao, Chao and Wang, Zhen and Kurths, Jurgen},
  journal={New Journal of Physics},
  volume={23},
  number={11},
  pages={113016},
  year={2021},
  publisher={IOP Publishing}
}

@article{Boguna:2009:Navigability:Networks,
   author = {Boguñá, Marián and Krioukov, Dmitri and Claffy, K. C.},
   title = {Navigability of complex networks},
   journal = {Nat. Phys.},
   volume = {5},
   number = {1},
   pages = {74-80},
   ISSN = {1745-2473
1745-2481},
   DOI = {10.1038/nphys1130},
   year = {2008},
   type = {Journal Article}
}

@article{Cheng:2010:Bridgeness:Edge:Significance,
   author = {Cheng, Xue-Qi and Ren, Fu-Xin and Shen, Hua-Wei and Zhang, Zi-Ke and Zhou, Tao},
   title = {Bridgeness: a local index on edge significance in maintaining global connectivity},
   journal = {J. Stat. Mech: Theory Exp.},
   volume = {2010},
   number = {10},
   ISSN = {1742-5468},
   DOI = {10.1088/1742-5468/2010/10/p10011},
   year = {2010},
   eid = {10011},
   type = {Journal Article}
}

@article{Cohen:2000:Resilience:Internet:Random:Breakdowns,
   author = {Cohen, Reuven and Erez, Keren and Ben-Avraham, Daniel and Havlin, Shlomo},
   title = {Resilience of the internet to random breakdowns},
   journal = {Phys. Rev. Lett.},
   volume = {85},
   number = {21},
   pages = {4626-4628},
   addendum = {Cohen, R
Erez, K
ben-Avraham, D
Havlin, S
eng
Phys Rev Lett. 2000 Nov 20;85(21):4626-8. doi: 10.1103/PhysRevLett.85.4626.},
   ISSN = {1079-7114 (Electronic)
0031-9007 (Linking)},
   DOI = {10.1103/PhysRevLett.85.4626},
   url = {https://www.ncbi.nlm.nih.gov/pubmed/11082612},
   year = {2000},
   type = {Journal Article}
}

@article{Cohen:2001:Breakdown:Internet:Intentional:Attack,
   author = {Cohen, Reuven and Erez, Keren and Ben-Avraham, Daniel and Havlin, Shlomo},
   title = {Breakdown of the internet under intentional attack},
   journal = {Phys. Rev. Lett.},
   volume = {86},
   number = {16},
   pages = {3682-3685},
   addendum = {Cohen, R
Erez, K
ben-Avraham, D
Havlin, S
eng
Research Support, U.S. Gov't, Non-P.H.S.
Phys Rev Lett. 2001 Apr 16;86(16):3682-5. doi: 10.1103/PhysRevLett.86.3682.},
   ISSN = {0031-9007 (Print)
0031-9007 (Linking)},
   DOI = {10.1103/PhysRevLett.86.3682},
   url = {https://www.ncbi.nlm.nih.gov/pubmed/11328053},
   year = {2001},
   type = {Journal Article}
}

@article{Gu:2021:Selection:Embedding:Dimension,
   author = {Gu, Weiwei and Tandon, Aditya and Ahn, Yong-Yeol and Radicchi, Filippo},
   title = {Principled approach to the selection of the embedding dimension of networks},
   journal = {Nat. Commun.},
   volume = {12},
   number = {1},
   ISSN = {2041-1723},
   DOI = {10.1038/s41467-021-23795-5},
   url = {https://doi.org/10.1038/s41467-021-23795-5},
   year = {2021},
   eid = {3772},
   type = {Journal Article}
}

@article{Kleineberg:2017:Geometric:Correlations:Mitigate:Vulnerabilities,
   author = {Kleineberg, Kaj-Kolja and Buzna, Lubos and Papadopoulos, Fragkiskos and Boguñá, Marián  and Serrano, M. {\'A}ngeles},
   title = {Geometric Correlations Mitigate the Extreme Vulnerability of Multiplex Networks against Targeted Attacks},
   journal = {Phys. Rev. Lett.},
   volume = {118},
   number = {21},
   addendum = {Kleineberg, Kaj-Kolja
Buzna, Lubos
Papadopoulos, Fragkiskos
Boguna, Marian
Serrano, M Angeles
eng
Phys Rev Lett. 2017 May 26;118(21):218301. doi: 10.1103/PhysRevLett.118.218301. Epub 2017 May 25.},
   ISSN = {1079-7114 (Electronic)
0031-9007 (Linking)},
   DOI = {10.1103/PhysRevLett.118.218301},
   url = {https://www.ncbi.nlm.nih.gov/pubmed/28598659},
   year = {2017},
   eid = {218301},
   type = {Journal Article}
}

@article{Kocc:2014:Impact:Cascading:Failures:Power:Grid,
   author = {Koç, Yakup and Warnier, Martijn and Mieghem, Piet Van and Kooij, Robert E. and Brazier, Frances M. T.},
   title = {The impact of the topology on cascading failures in a power grid model},
   journal = {Physica A},
   volume = {402},
   pages = {169-179},
   ISSN = {0378-4371},
   DOI = {https://doi.org/10.1016/j.physa.2014.01.056},
   url = {https://www.sciencedirect.com/science/article/pii/S0378437114000776},
   year = {2014},
   type = {Journal Article}
}

@article{Kojaku:2023:Community:Detection:Neural:Embedding,
   author = {Kojaku, Sadamori and Radicchi, Filippo and Ahn, Yong-Yeol and Fortunato, Santo},
   title = {Network Community Detection Via Neural Embeddings},
   journal = {Nature Communications},
   volume = {15},
   number = {1},
   pages = {9446},
   ISSN = {2041-1723},
   DOI = {10.1038/s41467-024-52355-w},
   url = {https://doi.org/10.1038/s41467-024-52355-w},
   year = {2024},
   type = {Journal Article}
}

@inproceedings{Macqueen:1967:K-Means,
   author = {MacQueen, James},
   title = {Some methods for classification and analysis of multivariate observations},
   booktitle = {Proceedings of the fifth Berkeley symposium on mathematical statistics and probability},
   publisher = {University of California Press},
   address = {Oakland, CA, USA},
   volume = {1},
   pages = {281-297},
   year = {1967},
   type = {Conference Proceedings}
}

@article{Makarov:2021:Survey:Graph:Embeddings:Applications,
   author = {Makarov, Ilya and Kiselev, Dmitrii and Nikitinsky, Nikita and Subelj, Lovro},
   title = {Survey on graph embeddings and their applications to machine learning problems on graphs},
   journal = {PeerJ Comput. Sci.},
   volume = {7},
   addendum = {Makarov, Ilya
Kiselev, Dmitrii
Nikitinsky, Nikita
Subelj, Lovro
eng
PeerJ Comput Sci. 2021 Feb 4;7:e357. doi: 10.7717/peerj-cs.357. eCollection 2021.},
   ISSN = {2376-5992 (Electronic)
2376-5992 (Linking)},
   DOI = {10.7717/peerj-cs.357},
   url = {https://www.ncbi.nlm.nih.gov/pubmed/33817007},
   year = {2021},
   eid = {e357},
   type = {Journal Article}
}

@article{Moriano:2019:Community:Event:Detection,
   author = {Moriano, Pablo and Finke, Jorge and Ahn, Yong-Yeol},
   title = {Community-Based Event Detection in Temporal Networks},
   journal = {Sci. Rep.},
   volume = {9},
   number = {1},
   ISSN = {2045-2322},
   DOI = {10.1038/s41598-019-40137-0},
   url = {https://doi.org/10.1038/s41598-019-40137-0},
   year = {2019},
   eid = {4358},
   type = {Journal Article}
}

@article{Osat:2023:Embedding:Aided:Dismantling,
   author = {Osat, Saeed and Papadopoulos, Fragkiskos and Teixeira, Andreia Sofia and Radicchi, Filippo},
   title = {Embedding-aided network dismantling},
   journal = {Phys. Rev. Res.},
   volume = {5},
   number = {1},
   ISSN = {2643-1564},
   DOI = {10.1103/PhysRevResearch.5.013076},
   year = {2023},
   eid = {013076},
   type = {Journal Article}
}

@article{Peng:2021:Neural:Embeddings:Periodicals,
   author = {Peng, Hao and Ke, Qing and Budak, Ceren and Romero, Daniel M and Ahn, Yong-Yeol},
   title = {Neural embeddings of scholarly periodicals reveal complex disciplinary organizations},
   journal = {Sci. Adv.},
   volume = {7},
   number = {17},
   addendum = {Peng, Hao
Ke, Qing
Budak, Ceren
Romero, Daniel M
Ahn, Yong-Yeol
eng
Research Support, Non-U.S. Gov't
Research Support, U.S. Gov't, Non-P.H.S.
Sci Adv. 2021 Apr 23;7(17):eabb9004. doi: 10.1126/sciadv.abb9004. Print 2021 Apr.},
   ISSN = {2375-2548 (Electronic)
2375-2548 (Linking)},
   DOI = {10.1126/sciadv.abb9004},
   url = {https://www.ncbi.nlm.nih.gov/pubmed/33893092},
   year = {2021},
   eid = {eabb9004},
   type = {Journal Article}
}

@article{Pereda:2019:Visualization:Graphs,
   author = {Pereda, María and Estrada, Ernesto},
   title = {Visualization and machine learning analysis of complex networks in hyperspherical space},
   journal = {Pattern Recognit.},
   volume = {86},
   pages = {320-331},
   ISSN = {0031-3203},
   DOI = {https://doi.org/10.1016/j.patcog.2018.09.018},
   url = {https://www.sciencedirect.com/science/article/pii/S0031320318303418},
   year = {2019},
   type = {Journal Article}
}

@inproceedings{Qiu-2018,
   author = {Qiu, Jiezhong and Dong, Yuxiao and Ma, Hao and Li, Jian and Wang, Kuansan and Tang, Jie},
   title = {Network embedding as matrix factorization: Unifying deepwalk, line, pte, and node2vec},
   booktitle = {Proceedings of the eleventh ACM international conference on web search and data mining},
   pages = {459-467},
   DOI = {10.1145/3159652.3159706},
   type = {Conference Proceedings},
   year = {2018}
}

@article{Rosenkrantz:2009:Resilience:Metrics:Service:Networks,
   author = {Rosenkrantz, Daniel J and Goel, Sanjay and Ravi, SS and Gangolly, Jagdish},
   title = {Resilience Metrics for Service-Oriented Networks: A Service Allocation Approach},
   journal = {IEEE Trans. Serv. Comput.},
   volume = {2},
   number = {3},
   pages = {183-196},
   ISSN = {1939-1374},
   DOI = {10.1109/TSC.2009.18},
   year = {2009},
   type = {Journal Article}
}

@article{Zeng:2012:Enhancing:Robustness:Malicious:Attacks,
   author = {Zeng, An and Liu, Weiping},
   title = {Enhancing network robustness against malicious attacks},
   journal = {Phys. Rev. E},
   volume = {85},
   number = {6 Pt 2},
   addendum = {Zeng, An
Liu, Weiping
eng
Research Support, Non-U.S. Gov't
Phys Rev E Stat Nonlin Soft Matter Phys. 2012 Jun;85(6 Pt 2):066130. doi: 10.1103/PhysRevE.85.066130. Epub 2012 Jun 27.},
   ISSN = {1550-2376 (Electronic)
1539-3755 (Linking)},
   DOI = {10.1103/PhysRevE.85.066130},
   url = {https://www.ncbi.nlm.nih.gov/pubmed/23005185},
   year = {2012},
   eid = {066130},
   type = {Journal Article}
}

@article{Albert-2000,
   author = {Albert, Réka and Jeong, Hawoong and Barabási, Albert-László},
   title = {Error and attack tolerance of complex networks},
   journal = {Nature},
   volume = {406},
   number = {6794},
   pages = {378-382},
   ISSN = {0028-0836},
   DOI = {10.1038/35019019},
   year = {2000},
   type = {Journal Article}
}

@article{Belkin-2003,
   author = {Belkin, Mikhail and Niyogi, Partha},
   title = {Laplacian Eigenmaps for Dimensionality Reduction and Data Representation},
   journal = {Neural Comput.},
   volume = {15},
   number = {6},
   pages = {1373-1396},
   ISSN = {0899-7667
1530-888X},
   DOI = {10.1162/089976603321780317},
   year = {2003},
   type = {Journal Article}
}

@article{Bellingeri-2020,
   author = {Bellingeri, Michele and Bevacqua, Daniele and Scotognella, Francesco and Alfieri, Roberto and Cassi, Davide},
   title = {A comparative analysis of link removal strategies in real complex weighted networks},
   journal = {Sci. Rep.},
   volume = {10},
   number = {1},
   ISSN = {2045-2322},
   DOI = {10.1038/s41598-020-60298-7},
   url = {https://doi.org/10.1038/s41598-020-60298-7},
   year = {2020},
   eid = {3911},
   type = {Journal Article}
}

@article{Boguna-2010,
   author = {Boguñá, Marián and Papadopoulos, Fragkiskos and Krioukov, Dmitri},
   title = {Sustaining the Internet with hyperbolic mapping},
   journal = {Nat. Commun.},
   volume = {1},
   ISSN = {2041-1723 (Electronic)
2041-1723 (Linking)},
   DOI = {10.1038/ncomms1063},
   url = {https://www.ncbi.nlm.nih.gov/pubmed/20842196},
   year = {2010},
   eid = {62},
   type = {Journal Article}
}

@article{BTWN,
   author = {Brandes, Ulrik},
   title = {A faster algorithm for betweenness centrality},
   journal = {J. Math. Sociol.},
   volume = {25},
   number = {2},
   pages = {163-177},
   ISSN = {0022-250X
1545-5874},
   DOI = {10.1080/0022250x.2001.9990249},
   year = {2001},
   type = {Journal Article}
}

@article{Clauset-2004,
   author = {Clauset, Aaron and Newman, Mark E. J. and Moore, Cristopher},
   title = {Finding community structure in very large networks},
   journal = {Phys. Rev. E},
   volume = {70},
   number = {6 Pt 2},
   ISSN = {1539-3755 (Print)
1539-3755 (Linking)},
   DOI = {10.1103/PhysRevE.70.066111},
   url = {https://www.ncbi.nlm.nih.gov/pubmed/15697438},
   year = {2004},
   eid = {066111},
   type = {Journal Article}
}

@article{Clauset-2009,
   author = {Clauset, Aaron and Shalizi, Cosma Rohilla and Newman, Mark E. J.},
   title = {Power-Law Distributions in Empirical Data},
   journal = {SIAM Rev.},
   volume = {51},
   number = {4},
   pages = {661-703},
   ISSN = {0036-1445
1095-7200},
   DOI = {10.1137/070710111},
   year = {2009},
   type = {Journal Article}
}

@article{Detection,
   author = {Fortunato, Santo},
   title = {Community detection in graphs},
   journal = {Phys. Rep.},
   volume = {486},
   number = {3-5},
   pages = {75-174},
   ISSN = {03701573},
   DOI = {10.1016/j.physrep.2009.11.002},
   year = {2010},
   type = {Journal Article}
}

@article{Fortunato-2016,
   author = {Fortunato, Santo and Hric, Darko},
   title = {Community detection in networks: A user guide},
   journal = {Phys. Rep.},
   volume = {659},
   pages = {1-44},
   ISSN = {0370-1573},
   DOI = {10.1016/j.physrep.2016.09.002},
   year = {2016},
   type = {Journal Article}
}

@inproceedings{NMI,
   author = {Fred, Ana Luisa Nobre and Jain, Anil K},
   title = {Robust data clustering},
   booktitle = {2003 IEEE Computer Society Conference on Computer Vision and Pattern Recognition, 2003. Proceedings.},
   publisher = {IEEE},
   address = {Madison, WI, USA},
   volume = {2},
   ISBN = {0769519008},
   DOI = {10.1109/CVPR.2003.1211462},
   year = {2003},
   type = {Conference Proceedings}
}

@article{clusim,
   author = {Gates, Alexander and Ahn, Yong-Yeol},
   title = {CluSim: a python package for calculating clustering similarity},
   journal = {J. Open Source Softw.},
   volume = {4},
   number = {35},
   ISSN = {2475-9066},
   DOI = {10.21105/joss.01264},
   year = {2019},
   eid = {01264},
   type = {Journal Article}
}

@article{ECS,
   author = {Gates, Alexander J and Wood, Ian B and Hetrick, William P and Ahn, Yong-Yeol},
   title = {Element-centric clustering comparison unifies overlaps and hierarchy},
   journal = {Sci. Rep.},
   volume = {9},
   number = {1},
   ISSN = {2045-2322},
   DOI = {10.1038/s41598-019-44892-y},
   year = {2019},
   eid = {8574},
   type = {Journal Article}
}

@article{GN,
   author = {Girvan, Michelle and Newman, Mark E. J.},
   title = {Community structure in social and biological networks},
   journal = {Proc. Natl. Acad. Sci. U.S.A.},
   volume = {99},
   number = {12},
   pages = {7821-7826},
   ISSN = {0027-8424 (Print)
1091-6490 (Electronic)
0027-8424 (Linking)},
   DOI = {10.1073/pnas.122653799},
   url = {https://www.ncbi.nlm.nih.gov/pubmed/12060727},
   year = {2002},
   type = {Journal Article}
}

@article{nxt-gem,
   author = {Goyal, Palash and Ferrara, Emilio},
   title = {GEM: a Python package for graph embedding methods},
   journal = {J. Open Source Softw.},
   volume = {3},
   number = {29},
   ISSN = {2475-9066},
   DOI = {10.21105/joss.00876},
   year = {2018},
   eid = {00876},
   type = {Journal Article}
}

@article{Goyal-2018,
   author = {Goyal, Palash and Ferrara, Emilio},
   title = {Graph embedding techniques, applications, and performance: A survey},
   journal = {Knowl.-Based Syst.},
   volume = {151},
   pages = {78--94},
   ISSN = {09507051},
   DOI = {10.1016/j.knosys.2018.03.022},
   year = {2018},
   type = {Journal Article}
}

@article{Guimera-2003,
   author = {Guimerà, Roger and Danon, Leon and Diaz-Guilera, Albert and Giralt, Francesc and Arenas, Alex},
   title = {Self-similar community structure in a network of human interactions},
   journal = {Phys. Rev. E},
   volume = {68},
   number = {6 Pt 2},
   ISSN = {1539-3755 (Print)
1539-3755 (Linking)},
   DOI = {10.1103/PhysRevE.68.065103},
   url = {https://www.ncbi.nlm.nih.gov/pubmed/14754250},
   year = {2003},
   eid = {065103},
   type = {Journal Article}
}

@techreport{nx,
   author = {Hagberg, Aric and Swart, Pieter and Schult, Daniel},
   title = {Exploring network structure, dynamics, and function using NetworkX},
   institution = {Los Alamos National Lab.(LANL), Los Alamos, NM (United States)},
   year = {2008},
   type = {Report}
}

@article{SB,
   author = {Holland, Paul William and Laskey, Kathryn Blackmond and Leinhardt, Samuel},
   title = {Stochastic blockmodels: First steps},
   journal = {Soc. Networks},
   volume = {5},
   number = {2},
   pages = {109-137},
   ISSN = {03788733},
   DOI = {10.1016/0378-8733(83)90021-7},
   year = {1983},
   type = {Journal Article}
}

@article{Karrer-2008,
   author = {Karrer, Brian and Levina, Elizaveta and Newman, Mark E. J.},
   title = {Robustness of community structure in networks},
   journal = {Phys. Rev. E},
   volume = {77},
   number = {4},
   DOI = {10.1103/PhysRevE.77.046119},
   year = {2008},
   eid = {046119},
   type = {Journal Article}
}

@article{LFR2,
   author = {Lancichinetti, Andrea and Fortunato, Santo},
   title = {Benchmarks for testing community detection algorithms on directed and weighted graphs with overlapping communities},
   journal = {Phys. Rev. E},
   volume = {80},
   number = {1 Pt 2},
   ISSN = {1539-3755 (Print)
1539-3755 (Linking)},
   DOI = {10.1103/PhysRevE.80.016118},
   url = {https://www.ncbi.nlm.nih.gov/pubmed/19658785},
   year = {2009},
   eid = {016118},
   type = {Journal Article}
}

@article{HHH,
   author = {Lancichinetti, Andrea and Fortunato, Santo},
   title = {Community detection algorithms: a comparative analysis},
   journal = {Phys. Rev. E},
   volume = {80},
   number = {5 Pt 2},
   ISSN = {1550-2376 (Electronic)
1539-3755 (Linking)},
   DOI = {10.1103/PhysRevE.80.056117},
   url = {https://www.ncbi.nlm.nih.gov/pubmed/20365053},
   year = {2009},
   eid = {056117},
   type = {Journal Article}
}

@article{LFR1,
   author = {Lancichinetti, Andrea and Fortunato, Santo and Radicchi, Filippo},
   title = {Benchmark graphs for testing community detection algorithms},
   journal = {Phys. Rev. E},
   volume = {78},
   number = {4 Pt 2},
   ISSN = {1539-3755 (Print)
1539-3755 (Linking)},
   DOI = {10.1103/PhysRevE.78.046110},
   url = {https://www.ncbi.nlm.nih.gov/pubmed/18999496
https://journals.aps.org/pre/pdf/10.1103/PhysRevE.78.046110},
   year = {2008},
   eid = {046110},
   type = {Journal Article}
}

@article{Jure-2007,
   author = {Leskovec, Jure and Kleinberg, Jon and Faloutsos, Christos},
   title = {Graph evolution: Densification and shrinking diameters},
   journal = {ACM Trans. Knowl. Discovery Data},
   volume = {1},
   number = {1},
   ISSN = {1556-4681},
   DOI = {10.1145/1217299.1217301},
   url = {https://doi.org/10.1145/1217299.1217301},
   year = {2007},
   eid = {2},
   type = {Journal Article}
}

@article{Lloyd-1982,
   author = {Lloyd, Stuart P.},
   title = {Least squares quantization in PCM},
   journal = {IEEE Trans. Inf. Theory},
   volume = {28},
   number = {2},
   pages = {129-137},
   ISSN = {0018-9448},
   DOI = {10.1109/tit.1982.1056489},
   year = {1982},
   type = {Journal Article}
}

@article{Mikolov-2013,
   author = {Mikolov, Tomas
Chen, Kai and Corrado, Greg and Dean, Jeffrey},
   title = {Efficient Estimation of Word Representations in Vector Space},
   year = {2013},
   journal={arXiv preprint arXiv:1301.3781},
   type = {Manuscript}
}

@inproceedings{Bayer-1999,
   author = {Beyer, Kevin and Goldstein, Jonathan and Ramakrishnan, Raghu and Shaft, Uri},
   title = {When is “nearest neighbor” meaningful?},
   booktitle = {Database Theory—ICDT’99: 7th International Conference Jerusalem, Israel, January 10–12, 1999 Proceedings 7},
   publisher = {Springer},
   address = {Berlin, Heidelberg},
   pages = {217-235},
   ISBN = {3540654526},
   year = {1999},
   type = {Conference Proceedings}
}

@inproceedings{Ou-2016,
   author = {Ou, Mingdong and Cui, Peng and Pei, Jian and Zhang, Ziwei and Zhu, Wenwu},
   title = {Asymmetric Transitivity Preserving Graph Embedding},
   booktitle = {KDD '16: Proceedings of the 22nd ACM SIGKDD International Conference on Knowledge Discovery and Data Mining},
   publisher = {Association for Computing Machinery},
   address = {New York, NY, USA},
   pages = {1105-1114},
   ISBN = {9781450342322},
   DOI = {10.1145/2939672.2939751},
   year = {2016},
   type = {Conference Proceedings}
}

@inproceedings{Liben:2003:Link:Prediction,
   author = {Liben-Nowell, David  and Kleinberg, Jon},
   title = {The link prediction problem for social networks},
   booktitle = {Proceedings of the twelfth international conference on Information and knowledge management},
   publisher = {Association for Computing Machinery},
   address = {New York, NY, USA},
   pages = {556--559},
   ISBN = {1581137230},
   DOI = {10.1145/956863.956972},
   year = {2003},
   type = {Conference Proceedings}
}

@article{Palla-2005,
   author = {Palla, Gergely and Derenyi, Imre and Farkas, Illés and Vicsek, Tamás},
   title = {Uncovering the overlapping community structure of complex networks in nature and society},
   journal = {Nature},
   volume = {435},
   number = {7043},
   pages = {814-818},
   ISSN = {1476-4687 (Electronic)
0028-0836 (Linking)},
   DOI = {10.1038/nature03607},
   url = {https://www.ncbi.nlm.nih.gov/pubmed/15944704},
   year = {2005},
   type = {Journal Article}
}

@inproceedings{Perozzi-2014,
   author = {Perozzi, Bryan and Al-Rfou, Rami and Skiena, Steven},
   title = {Deepwalk: Online learning of social representations},
   booktitle = {Proceedings of the 20th ACM SIGKDD international conference on Knowledge discovery and data mining},
   pages = {701-710},
   DOI = {10.1145/2623330.2623732},
   year = {2014},
   type = {Conference Proceedings}
}

@article{Reynolds-2009,
   author = {Reynolds, Douglas A.},
   title = {Gaussian mixture models},
   journal = {Encyclopedia of biometrics},
   volume = {741},
   number = {659-663},
   pages = {827–832},
   year = {2009},
   type = {Journal Article}
}

@article{Sam-2000,
   author = {Roweis, Sam T. and Saul, Lawrence K.},
   title = {Nonlinear dimensionality reduction by locally linear embedding},
   journal = {Science},
   volume = {290},
   number = {5500},
   pages = {2323-2326},
   ISSN = {0036-8075 (Print)
0036-8075 (Linking)},
   DOI = {10.1126/science.290.5500.2323},
   url = {https://www.ncbi.nlm.nih.gov/pubmed/11125150},
   year = {2000},
   type = {Journal Article}
}

@inproceedings{karate,
   author = {Rozemberczki, Benedek and Kiss, Oliver and Sarkar, Rik},
   title = {Karate Club: an API oriented open-source python framework for unsupervised learning on graphs},
   booktitle = {Proceedings of the 29th ACM international conference on information and knowledge management},
   pages = {3125-3132},
   DOI = {10.1145/3340531.3412757},
   year = {2020},
   type = {Conference Proceedings}
}

@inproceedings{Schubert-2021,
   author = {Schubert, Erich and Lang, Andreas and Feher, Gloria},
   title = {Accelerating spherical k-means},
   booktitle = {International Conference on Similarity Search and Applications},
   publisher = {Springer},
   address = {Vienna, Austria},
   pages = {217-231},
   year = {2021},
   type = {Conference Proceedings}
}

@article{Tandon-2021,
   author = {Tandon, Aditya and Albeshri, Aiiad and Thayananthan, Vijey and Alhalabi, Wadee and Radicchi, Filippo and Fortunato, Santo},
   title = {Community detection in networks using graph embeddings},
   journal = {Phys. Rev. E},
   volume = {103},
   number = {2-1},
   ISSN = {2470-0053 (Electronic)
2470-0045 (Linking)},
   DOI = {10.1103/PhysRevE.103.022316},
   url = {https://www.ncbi.nlm.nih.gov/pubmed/33736102
https://journals.aps.org/pre/abstract/10.1103/PhysRevE.103.022316},
   year = {2021},
   eid = {022316},
   type = {Journal Article}
}

@article{Tian-2023,
   author = {Tian, Moyi and Moriano, Pablo},
   title = {Robustness of community structure under edge addition},
   journal = {Phys. Rev. E},
   volume = {108},
   number = {5},
   addendum = {PRE},
   DOI = {10.1103/PhysRevE.108.054302},
   url = {https://link.aps.org/doi/10.1103/PhysRevE.108.054302},
   year = {2023},
   eid = {054302},
   type = {Journal Article}
}

@article{LUX,
   author = {von Luxburg, Ulrike},
   title = {A tutorial on spectral clustering},
   journal = {Stat. Comput.},
   volume = {17},
   number = {4},
   pages = {395-416},
   ISSN = {0960-3174
1573-1375},
   DOI = {10.1007/s11222-007-9033-z},
   year = {2007},
   type = {Journal Article}
}

@article{Wang-2018,
   author = {Wang, Shuai and Liu, Jing},
   title = {Constructing robust community structure against edge-based attacks},
   journal = {IEEE Syst. J.},
   volume = {13},
   number = {1},
   pages = {582-592},
   ISSN = {1932-8184},
   DOI = {10.1109/JSYST.2018.2835642},
   year = {2018},
   type = {Journal Article}
}

@article{Wang-2017,
   author = {Wang, Shuai and Liu, Jing and Wang, Xiaodong},
   title = {Mitigation of attacks and errors on community structure in complex networks},
   journal = {J. Stat. Mech: Theory Exp.},
   volume = {2017},
   number = {4},
   ISSN = {1742-5468},
   DOI = {10.1088/1742-5468/aa6581},
   year = {2017},
   eid = {043405},
   type = {Journal Article}
}

@inproceedings{WangX-2017,
   author = {Wang, Xiao and Cui, Peng and Wang, Jing and Pei, Jian and Zhu, Wenwu and Yang, Shiqiang},
   title = {Community Preserving Network Embedding},
   booktitle = {Proc. AAAI Conf. Artif. Intell.},
   volume = {31},
   year = {2017},
   ISBN = {2374-3468
2159-5399},
   DOI = {10.1609/aaai.v31i1.10488},
   type = {Conference Proceedings}
}

@article{Xu-2021,
   author = {Xu, Mengjia},
   title = {Understanding Graph Embedding Methods and Their Applications},
   journal = {SIAM Rev.},
   volume = {63},
   number = {4},
   pages = {825-853},
   ISSN = {0036-1445
1095-7200},
   DOI = {10.1137/20m1386062},
   year = {2021},
   type = {Journal Article}
}

@article{Yan-2007,
   author = {Yan, Shuicheng and Xu, Dong and Zhang, Benyu and Zhang, Hong-Jiang and Yang, Qiang and Lin, Stephen},
   title = {Graph Embedding and Extensions: A General Framework for Dimensionality Reduction},
   journal = {IEEE Trans. Pattern Anal. Mach. Intell.},
   volume = {29},
   number = {1},
   pages = {40-51},
   ISSN = {0162-8828
2160-9292},
   DOI = {10.1109/tpami.2007.250598},
   year = {2007},
   type = {Journal Article}
}

@article{LS1-1,
  title={Iterative embedding and reweighting of complex networks reveals community structure},
  author={Kov{\'a}cs, Bianka and Kojaku, Sadamori and Palla, Gergely and Fortunato, Santo},
  journal={Scientific Reports},
  volume={14},
  number={1},
  pages={17184},
  year={2024},
  publisher={Nature Publishing Group UK London}
}

@article{LS1-4,
author = {Jingnan Zhang, Xin He and Junhui Wang},
title = {Directed Community Detection With Network Embedding},
journal = {Journal of the American Statistical Association},
volume = {117},
number = {540},
pages = {1809--1819},
year = {2022},
publisher = {ASA Website},
doi = {10.1080/01621459.2021.1887742},
URL = {https://doi.org/10.1080/01621459.2021.1887742},
eprint = {https://doi.org/10.1080/01621459.2021.1887742}
}

@article{LS1-5,
    author = {Pankratz, Bartosz and Kamiński, Bogumił and Prałat, Paweł},
    title = {Performance of community detection algorithms supported by node embeddings},
    journal = {Journal of Complex Networks},
    volume = {12},
    number = {4},
    pages = {cnae035},
    year = {2024},
    month = {08},
    issn = {2051-1329},
    doi = {10.1093/comnet/cnae035},
    url = {https://doi.org/10.1093/comnet/cnae035},
    eprint = {https://academic.oup.com/comnet/article-pdf/12/4/cnae035/60799757/cnae035.pdf},
}

@INPROCEEDINGS{LS1-6,
  author={Ghanbari, Mahdi and Chehreghani, Mostafa Haghir and Chehreghani, Morteza Haghir},
  booktitle={2022 IEEE International Conference on Big Data (Big Data)}, 
  title={Graph Clustering Using Node Embeddings: An Empirical Study}, 
  year={2022},
  volume={},
  number={},
  pages={5488-5493},
  doi={10.1109/BigData55660.2022.10020377}
}

@article{LS2-1,
  title={Robustness in network community detection under links weights uncertainties},
  author={Ramirez-Marquez, Jose Emmanuel and Rocco, CM and Moronta, Jos{\'e} and Dessavre, D Gama},
  journal={Reliability Engineering \& System Safety},
  volume={153},
  pages={88--95},
  year={2016},
  publisher={Elsevier}
}

@inproceedings{LS2-2,
 author = {Peche, Sandrine and Perchet, Vianney},
 booktitle = {Advances in Neural Information Processing Systems},
 editor = {H. Larochelle and M. Ranzato and R. Hadsell and M.F. Balcan and H. Lin},
 pages = {17827--17837},
 publisher = {Curran Associates, Inc.},
 address = {Red Hook, NY, USA},
 title = {Robustness of Community Detection to Random Geometric Perturbations},
 volume = {33},
 year = {2020}
}

@inproceedings{LS2-3,
author = {Jia, Jinyuan and Wang, Binghui and Cao, Xiaoyu and Gong, Neil Zhenqiang},
title = {Certified Robustness of Community Detection against Adversarial Structural Perturbation via Randomized Smoothing},
year = {2020},
isbn = {9781450370233},
publisher = {Association for Computing Machinery},
address = {New York, NY, USA},
url = {https://doi.org/10.1145/3366423.3380029},
doi = {10.1145/3366423.3380029},
booktitle = {Proceedings of The Web Conference 2020},
pages = {2718–2724},
numpages = {7},
location = {Taipei, Taiwan},
series = {WWW '20}
}

@InProceedings{zhu-2021,
author="Zhang, Fan
and Zhu, Junyou
and Luo, Zheng
and Wang, Zhen
and Tao, Li
and Gao, Chao",
editor="Qiu, Han
and Zhang, Cheng
and Fei, Zongming
and Qiu, Meikang
and Kung, Sun-Yuan",
title="Community Detection in Dynamic Networks: A Novel Deep Learning Method",
booktitle="Knowledge Science, Engineering and Management",
year="2021",
publisher="Springer International Publishing",
address="Cham",
pages="115--127",
isbn="978-3-030-82136-4"
}

@ARTICLE{sun-2022,
  author={Sun, Jianyong and Zheng, Wei and Zhang, Qingfu and Xu, Zongben},
  journal={IEEE Transactions on Cybernetics}, 
  title={Graph Neural Network Encoding for Community Detection in Attribute Networks}, 
  year={2022},
  volume={52},
  number={8},
  pages={7791-7804},
  doi={10.1109/TCYB.2021.3051021}
}

@ARTICLE{su-2024,
  author={Su, Xing and Xue, Shan and Liu, Fanzhen and Wu, Jia and Yang, Jian and Zhou, Chuan and Hu, Wenbin and Paris, Cecile and Nepal, Surya and Jin, Di and Sheng, Quan Z. and Yu, Philip S.},
  journal={IEEE Transactions on Neural Networks and Learning Systems}, 
  title={A Comprehensive Survey on Community Detection With Deep Learning}, 
  year={2024},
  volume={35},
  number={4},
  pages={4682-4702},
  doi={10.1109/TNNLS.2021.3137396}}

@article{JMLR:v24:20-998,
  author  = {Anton Tsitsulin and John Palowitch and Bryan Perozzi and Emmanuel Müller},
  title   = {Graph Clustering with Graph Neural Networks},
  journal = {Journal of Machine Learning Research},
  year    = {2023},
  volume  = {24},
  number  = {127},
  pages   = {1--21},
  url     = {http://jmlr.org/papers/v24/20-998.html}
}

@inproceedings{NEURIPS2021_ca954182,
 author = {Kojaku, Sadamori and Yoon, Jisung and Constantino, Isabel and Ahn, Yong-Yeol},
 booktitle = {Advances in Neural Information Processing Systems},
 editor = {M. Ranzato and A. Beygelzimer and Y. Dauphin and P.S. Liang and J. Wortman Vaughan},
 pages = {24150--24163},
 publisher = {Curran Associates, Inc.},
 address = {Red Hook, NY, USA},
 title = {Residual2Vec: Debiasing graph embedding with random graphs},
 volume = {34},
 year = {2021}
}

@article{PhysRevLett.85.4626,
  title = {Resilience of the Internet to Random Breakdowns},
  author = {Cohen, Reuven and Erez, Keren and ben-Avraham, Daniel and Havlin, Shlomo},
  journal = {Phys. Rev. Lett.},
  volume = {85},
  issue = {21},
  pages = {4626--4628},
  numpages = {0},
  year = {2000},
  month = {Nov},
  publisher = {American Physical Society},
  doi = {10.1103/PhysRevLett.85.4626},
  url = {https://link.aps.org/doi/10.1103/PhysRevLett.85.4626}
}

@article{robert,
title = {Tails of random sums of a heavy-tailed number of light-tailed terms},
journal = {Insurance: Mathematics and Economics},
volume = {43},
number = {1},
pages = {85-92},
year = {2008},
issn = {0167-6687},
doi = {https://doi.org/10.1016/j.insmatheco.2007.10.001},
url = {https://www.sciencedirect.com/science/article/pii/S0167668707001084},
author = {Christian Y. Robert and Johan Segers}
}

@inproceedings{Grover-2016,
author = {Grover, Aditya and Leskovec, Jure},
title = {node2vec: Scalable Feature Learning for Networks},
year = {2016},
isbn = {9781450342322},
publisher = {Association for Computing Machinery},
address = {New York, NY, USA},
url = {https://doi.org/10.1145/2939672.2939754},
doi = {10.1145/2939672.2939754},
booktitle = {Proceedings of the 22nd ACM SIGKDD International Conference on Knowledge Discovery and Data Mining},
pages = {855–864},
numpages = {10},
location = {San Francisco, California, USA},
series = {KDD '16}
}

@inproceedings{Tang-2015,
author = {Tang, Jian and Qu, Meng and Wang, Mingzhe and Zhang, Ming and Yan, Jun and Mei, Qiaozhu},
title = {LINE: Large-scale Information Network Embedding},
year = {2015},
isbn = {9781450334693},
publisher = {International World Wide Web Conferences Steering Committee},
address = {Republic and Canton of Geneva, CHE},
url = {https://doi.org/10.1145/2736277.2741093},
doi = {10.1145/2736277.2741093},
booktitle = {Proceedings of the 24th International Conference on World Wide Web},
pages = {1067–1077},
numpages = {11},
location = {Florence, Italy},
series = {WWW '15}
}

@article{Cannistraci:2013:Link,
  title={From link-prediction in brain connectomes and protein interactomes to the local-community-paradigm in complex networks},
  author={Cannistraci, Carlo Vittorio and Alanis-Lobato, Gregorio and Ravasi, Timothy},
  journal={Scientific reports},
  volume={3},
  number={1},
  pages={1613},
  year={2013},
  publisher={Nature Publishing Group UK London}
}

\end{document}